\documentclass{PoSarxiv}

\title{Stochastic Motion of Heavy Quarks in Holography: A Theory-Independent Treatment}

\ShortTitle{Heavy Quark Noise in Holography}

\author{Dimitrios Giataganas\\
        Physics Division, National Center for Theoretical
   Sciences, \\
  National Tsing-Hua University, Hsinchu, 30013, Taiwan\\
        E-mail: \email{dimitrios.giataganas@cts.nthu.edu.tw}}

\abstract{Stochastic dynamics play a central role in strongly coupled phenomena. We present and review a theory independent approach in holography to study such phenomena. We firstly argue that the heavy quark diffusion occurs in realistic strongly coupled systems. Then we analyze the quantum and thermal fluctuation, dissipation  and the corresponding Brownian motion of a heavy particle in such environments for a wide class of theories. The holographic study is based on the properties of the straight string fluctuations. The observables and coefficients associated with the stochastic motion depend on a single parameter which encodes the properties of the different theories. Moreover, certain Dp-brane fluctuations can be mapped one-to-one to the string fluctuations and therefore the stochastic brane observables can be read from the string ones.

Then we review the Langevin diffusion of a moving heavy quark in generic thermal holographic theories. The analysis is based on the properties of the trailing string and its fluctuations. The string world-sheet has a black hole horizon and the quark feels an effective temperature different than the environmental one. The formulas of the effective temperature, the drag force on the particle and the Langevin coefficients are given in terms of the background metric elements readily applicable to any theory. At the end we comment on the backreaction effects on the medium and present results of the Monte Carlo simulations.}

\FullConference{~}

\begin{document}

\section{Brief introduction to Langevin Dynamics}

Non-equilibrium systems evolving towards equilibrium exist everywhere in nature and have a wide range of applications, from soft matter and biophysics to strongly coupled phenomena in the most extreme phases of matter. The point of converging of such phenomena is the Brownian motion \cite{brown01}, the irregular motion of mesoscopic particles that are large enough to admit a hydrodynamic type of coarse-graining, but small enough to exhibit thermal fluctuations in a liquid environment caused by random microscopic interactions with the particles of the medium. The implications of Brownian motion were very important even from the initial stages of their theoretical study, verifying  at the time the existence of atomic nature of matter and providing a good estimation of the Avogadro's constant \cite{ein1905,sutherland1, smol1,perrin1}. Shortly after these developments, the probabilistic description was introduced to lay down the foundation of non-equilibrium statistical mechanics with the Langevin and Fokker-Planck equations. Since then there is a continuous interest on theoretical developments and new applications in all quantitative sciences, making the topic, a slow evolving revolution \cite{brownianII,brownian100,brownianrel,brownian111}.

The fluctuation and dissipation effects can be realized in the classical Langevin equation
\bea \label{lang}
\dot{p}_i(t)=-\eta_D p_i(t) +\xi_i(t)~,
\eea
where $p$ is the momentum of the particle, $\eta_D$ is the so called friction coefficient or the momentum drag coefficient and $\xi$ is the random force. These two competing phenomena are realized in nature by the friction force corresponding to dissipation, representing the energy transfer to the environment, and the noise of the environment which contributes the right amount of energy to the system through the fluctuations so that it evolves towards equilibrium. The physical picture described can be formulated as a theorem: the fluctuation-dissipation theorem relating the magnitudes of these phenomena while at equilibrium.

The theoretical developments are based on natural assumptions, mainly on separation of the timescales of the phenomena involved. These are the larger relaxation time needed for the particle to thermalize compared to  the smaller collision timescale. The random force then has the following properties: i) for timescales larger the collision time the force is stochastic with zero mean value, ii) the stochastic force is uncorrelated to itself and is translationally invariant, property triggered by the equilibrated homogeneous environment, iii) the statistical properties of the random force obey the time-translational invariance. These are  summarized to the following relations
\bea
\vev{\xi(t)}=0~,\qquad \vev{\xi(t)\xi(t')}=\kappa \delta(t-t')~,
\eea
where $\kappa$ is a constant measuring the degree of correlation and represents the mean squared momentum per unit of time. A particle experienced such a random force undergoes a random independent displacement generated by the integral
\bea
\int_0^t \xi(t')dt'= \int_0^{t_1} \xi(t')dt'+\int_{t_1}^{t_2} \xi(t')dt'+\ldots~,
\eea
which can be thought as a summation of independent terms, each one drawn from the same distribution, resulting to a total integral obeying a normal distribution with zero mean. By applying the central limit theorem of statistics one may obtain directly $\left< \dx^2 \right>\sim t$. Alternatively, the solution of the stochastic equation for $\tau\gg \eta_D^{-1}$ is
\bea\label{p2}
p_i(t)=\int_{-\infty}^t dt' e^{\eta_D(t'-t)}\xi_i(t')~,\qquad \vev{p^2}=\int^t dt_1 dt_2  e^{\eta_D(t_1+t_2)} \vev{\xi_i(t_1)\xi_i(t_2)}~=\frac{3\kappa}{2 \eta_D}~.
\eea
From the equipartition theorem the typical thermal momentum reads $p\sim \sqrt{M T}$ and therefore the drag coefficient is related to temperature \eq{p2} as
\bea\label{eta}
\eta_D=\frac{\kappa}{2 M T}~.
\eea
To introduce the diffusion coefficient we compute the mean squared position of the particle from \eq{lang} at later time and by using again the  equipartition theorem we get
\bea
\vev{x_i(t)x_j(t)}=2 D t \delta_{ij}~, \qquad \vev{x^2(t)}=\frac{1}{M^2} \int^t dt_1 dt_2\vev{p(t_1)p(t_2)}=  \frac{6 T t}{M \eta_D}~,
\eea
relating the diffusion constant with respect to the drag coefficient \eq{eta} and to the mean squared momentum transfer as
\bea\label{eqdn}
D=\frac{T}{M\eta_D}=\frac{2 T^2}{\kappa}~.
\eea

\subsection{Justification of the Heavy Quark Diffusion in the Quark-Gluon Plasma}

The task of modeling the heavy quark interaction in thermal environment is amenable to similar diffusion treatment we have just described \cite{Svetitsky:1987gq,vanHees:2004gq,Moore:2004tg,Mustafa:2004dr}. The role of the heavy test particle of the previous paragraph undergoing a Brownian motion, is played by a heavy quark in an environment of a light-particle fluid. The charm and bottom quark masses are much larger than the temperature and the constituent masses of the equilibrated Quark-Gluon-Plasma (QGP) environment, providing a good separation between the relaxation and the collision times.
Let us justify this by the following estimations.

The typical non-relativistic thermal momentum of a heavy quark with mass M with $M\gg T$, is $p^2\sim M T\gg T^2$ resulting to the low velocity $v^2\sim T/M\ll 1$. The typical square momentum transfer from the medium for hard collisions is of order $Q^2\sim T^2$ from the equipartition theorem, therefore we eventually have $p\gg Q$. As a result a large number of collisions of order $M/T$ is required to change the momentum by a factor of order one. Therefore, the interaction of the heavy quark with the medium can be formulated with uncorrelated momentum kicks, and the Boltzmann equation in momentum transfer can be expanded to the Fokker-Planck equation of heavy quark diffusion, in the medium of the QGP. We can further estimate $\a_S\simeq 0.5$ and $M/T=7$ to obtain the drag coefficient with respect to the diffusion constant $\eta_D^{-1}\simeq 7 D$.

Therefore, the Brownian motion and the dissipation phenomena provide direct observables for the heavy quarks in strongly coupled theories, capable to reveal potential interesting properties of the fundamental interactions. In a sense the heavy quark interactions can be thought as encoded in the transport coefficients, which in principle are related to the scattering matrix elements on light partons in the QGP.

One way to study the heavy quark diffusion is by perturbation theory \cite{Svetitsky:1987gq}, where the medium is approximated as a weakly interacting system of quark and gluons, treatment that is not reliable in conditions realized at heavy ion colliders. Non-perturbative interactions can be captured by effective resonance models \cite{vanHees:2004gq} allowing to compute certain resonances related to the diffusion dynamics, or by methods of holography which we discuss in this review.

\section{Fluctuation and Dissipation in Holography: A Unified Approach}

In this section we focus mostly on the description of heavy quark diffusion in the context of the gauge/gravity duality
reviewing mainly the unified study scheme developed in \cite{Giataganas:2018ekx}. Previous works initializing the ideas on AdS and Lifshitz spacetime include \cite{deBoer:2008gu,Son:2009vu,CaronHuot:2011dr,Sonner:2012if,Tong:2012nf}. While other relevant works include \cite{Hubeny:2010ry,Fischler:2014tka,Yeh:2014mfa,Yeh:2015cra,Banerjee:2015vmo,Moerman:2016wpv,Lee:2016wcn, Kiritsis:2013iba, Ho:2013rra,Fischler:2012ff,Roychowdhury:2015mta}. In the next section we will study the fluctuations and the energy loss of a moving heavy quark.

To present the general picture, let us introduce the gravity dual theory  in string frame
\bea\label{gen1}
ds^2=-g_{00}(r) dx_0^2+ g_{rr}(r)dr^2 +\sum_{i=1}^{d} g_{ii}(r) dx_i^2 ~,
\eea
with $\lim_{r\to \infty} g_{ii}(r)=\infty$, such that the boundary is at $r=\infty$. $d$ are the space dimensions and the metric is diagonal. The dual field theory lives on the spacetime spanned by $(x_0,x_i)$ and $r$ is the holographic direction.

The massive  heavy particle is represented by a string initiating from to the boundary spacetime, introducing therefore to the field theory extra degrees of freedom, and extending to the bulk of the space in the IR until the point $r= r_h$. For a thermal quantum field theory the $r_h\neq 0$ corresponding to the horizon of the black hole, while for a field theory at zero temperature  $r_h=0$ and the string terminates at the deep IR. The dynamics of such strings are described by solutions of the  Nambu-Goto (NG) action. Let us consider a worldsheet which extends along the $x_1$ direction being parametrized by $x_1=x_1(\tau,\sigma),~ u=\sigma,~ x_0=\tau,$ where $(\tau,\sigma)$ are the worldsheet coordinates. The action is equal to
\bea\label{actiona1}
S=-\frac{1}{2 \pi \alpha'} \int d\sigma d\tau \sqrt{-\prt{g_{00}+g_{11} \dx_1^2} \prt{g_{rr}+g_{11} x_1'^2}}~.
\eea
For a static particle one expects by symmetry arguments a straight string solution. Indeed it can be easily found that the solution to the equations of motion of the above action is $x_1=0$, where a space coordinate transformation is implemented to bring the origin at the position of the string.

The fluctuations of the heavy quark are realized by the dynamics of string fluctuations. For the static quark we need to consider the fluctuations around $x_1=0$. However, the length of the string-worldsheet in the bulk is infinite due to the infinite distance of the boundary of the space to its interior. Since the length is proportional to the particle's mass, we need to introduce a Neumann boundary condition on the location of a flavor brane $r_b$ close to the boundary. The boundary condition then reads $x_1'(r_b)=0$. The fluctuations $\delta x_1(t,r)$ give the Nambu-Goto action in terms of the metric elements \cite{Giataganas:2018ekx}
\bea\label{oactionorder2}
S=~c~-\frac{1}{4 \pi \alpha'} \int d\sigma d\tau  \left(-\frac{g_{11} \sqrt{-g_{00}}}{\sqrt{g_{rr}}}\delta x_1'^2+ \frac{g_{11}\sqrt{g_{rr}}}{\sqrt{-g_{00}}}\delta \dx_1^2\right)~.
\eea
The Fourier decomposed fluctuations then take the form
\bea\label{flucgen}
\delta x_1(t,r)=\int_0^\infty h_\omega (r) \left(\alpha(\omega)e^{-i\omega \tau}+\alpha(\omega)^\dagger e^{i\omega \tau}\right)~,
\eea
with $\alpha(\omega)^\dagger$ and  $ \alpha(\omega)$ being the creation and annihilation operators, and the mode equation reads
\bea\label{modes0}
\frac{\pp}{\pp r}\prt{\frac{g_{11} \sqrt{-g_{00}}}{g_{rr}}h_\omega(r)'}+\omega^2\frac{g_{11}\sqrt{g_{rr}}}{\sqrt{g_{00}}}h_{\omega}(r)=0~.
\eea

\subsection{Heavy Quark Fluctuation at Zero Temperature}

It is known that in  quantum physics the fluctuation and dissipation phenomena are present  even at zero temperature due to vacuum fluctuations of the environment fields and the uncertainty principle. Even in the simplest case of the zero-point energy of the electromagnetic field it has been shown that the fluctuation and dissipation phenomena occur and the relevant theorems hold \cite{landausp}. Perturbative methods analyzing these phenomena have also been developed to integrate out the environmental degrees of freedom by modeling them as an infinite number of  simple harmonic oscillators \cite{caldeira1983,Schwinger,Feynman:1963fq,Grabert:1988yt, Hu:1993qa, Hu:1986jj}. Ohmic, sub-subohmic, supra-ohmic environmental effects have been analyzed in such approaches, for example works on the latter ones include \cite{Hsiang:2005pz, Hsiang:2007zb}.

The quantum fluctuations on the heavy quark in a strongly coupled environment follows the same logic. The test particle fluctuations are induced by its coupling to the gluonic fields, resulting to a non-uniform motion. The dissipation is realized by a gluonic radiation back to the medium induced by the non-canonical motion of the quark. Here we analyze the resummed physical effects of such theories using techniques of gauge/gravity duality, reviewing the generical methodology developed in \cite{Giataganas:2018ekx}.

To proceed to the solution of the fluctuation equations, it is necessary to consider a certain general class of theories with dual backgrounds that belong to \eq{gen1}, which we choose as
\bea\label{polymetric1}
g_{00}=r^{a_0} f(r)~,\quad g_{rr}=\frac{1}{r^{a_u} f(r)}~,\quad g_{ii}=r^{a_i}~,\quad f(r)=1~,\quad r_h=0~,
\eea
where the $i$ indices count the spatial directions and $a_i$ are constant powers. The class of the dual field theories accommodated by the above metric includes the hyperscaling Lifshitz violating ones \cite{Kachru:2008yh,Dong:2012se,Narayan:2012hk}, the anisotropic theories \cite{Azeyanagi:2009pr,Mateos:2011ix,Mateos:2011tv,Giataganas:2017koz,Jain:2014vka,Donos:2016zpf} and several other. Features of the current analysis are applicable for  backgrounds with asymptotics of \eq{polymetric1} like certain RG flow gravity dual solutions.

There are several reasons for choosing the form of the metric as in \eq{polymetric1}. Certain rescaling can bring it to a form with less number of constants $a_i$, however this would not make our presentation simpler, since our results are formulated in terms of a constant $\nu$ incorporating all the scalings of the background. Moreover, formulas derived below with the form of the chosen metric are directly applicable to any gravity background accommodated, without the need of any coordinate transformation. Finally, such choice is convenient to build on the finite temperature string fluctuation analysis and to make the mapping of the string-brane fluctuations we study in later sections.

Notice that the string fluctuations in hyperscaling violation theories at zero temperature \cite{Edalati:2012tc}  guarantees that our methods at zero temperature go through all the way for particular metrics although the theories considered here are more generic than the hyperscaling ones. These include for example the anisotropic theories with different stability and physical ranges of the parameters, compared to hyperscaling theories, allowing new features on string fluctuations. Our notation offers also a powerful advantage, since each scaling is unique, we track accurately how the different metric elements affect the observables. This is the crucial point that will allows us to identify the order of the Bessel function of the fluctuations as the central quantity that the stochastic observables depend exclusively on. This observation holds even in finite temperature as we will show in later sections.

For this class of the background, the mode equations \eq{modes0} becomes
\bea\label{mawtw1}
\frac{\partial}{\partial r}\left(r^{a_1+\frac{a_0+a_u}{2}} h_\omega(r)'\right)+\omega^2 r^{a_1-\frac{a_0+a_u}{2}}h_{\omega}(r) = 0 ~ ,
\eea
with a solution of type \cite{Giataganas:2018ekx}
\bea\label{solutionmodes1}
h_\omega(r) = r^{-\nu \kappa}  A_\omega \prtt{J_\nu\prt{\omega \tilde{r}}+B_\omega Y_\nu\prt{\omega \tilde{r}}}~,
\eea
where $A_\omega$ and $B_\omega$ are constants and
\bea\label{definitionn1}
\tilde{r} := \frac{2 r^{\frac{1}{2}\left(2-\a_0-\a_u\right)}}{a_0+a_u-2}=\frac{r^{-\kappa}}{\kappa}~,\qquad \kappa:= \frac{a_0+a_u}{2}-1~,\qquad \nu:=\frac{a_0+2 a_1+a_u-2}{2\prt{a_0+a_u-2}}~
\eea
and $J_\nu(\tilde{r}),~Y_\nu(\tilde{r})$ are the Bessel functions of first and second kind. The integration constants are found by looking at the canonical computation relations for theories in curved space-times and at the Neumann boundary condition to obtain (Figure 1)
\bea
A_\om=\sqrt{\frac{  \pi \a'}{|\kappa|\prt{1+B_\omega^2}}}~, \qquad B_\omega=- \frac{J_{\n-1}\prt{\omega \tilde{r}_b}}{Y_{\n-1}\prt{ \omega\trr_b}}~,
\eea
where in the above and following computations we need to use the Bessel function properties presented in the Appendix  of  \cite{Giataganas:2018ekx}.

\begin{figure}
\centerline{\includegraphics[width=60mm,angle =90]{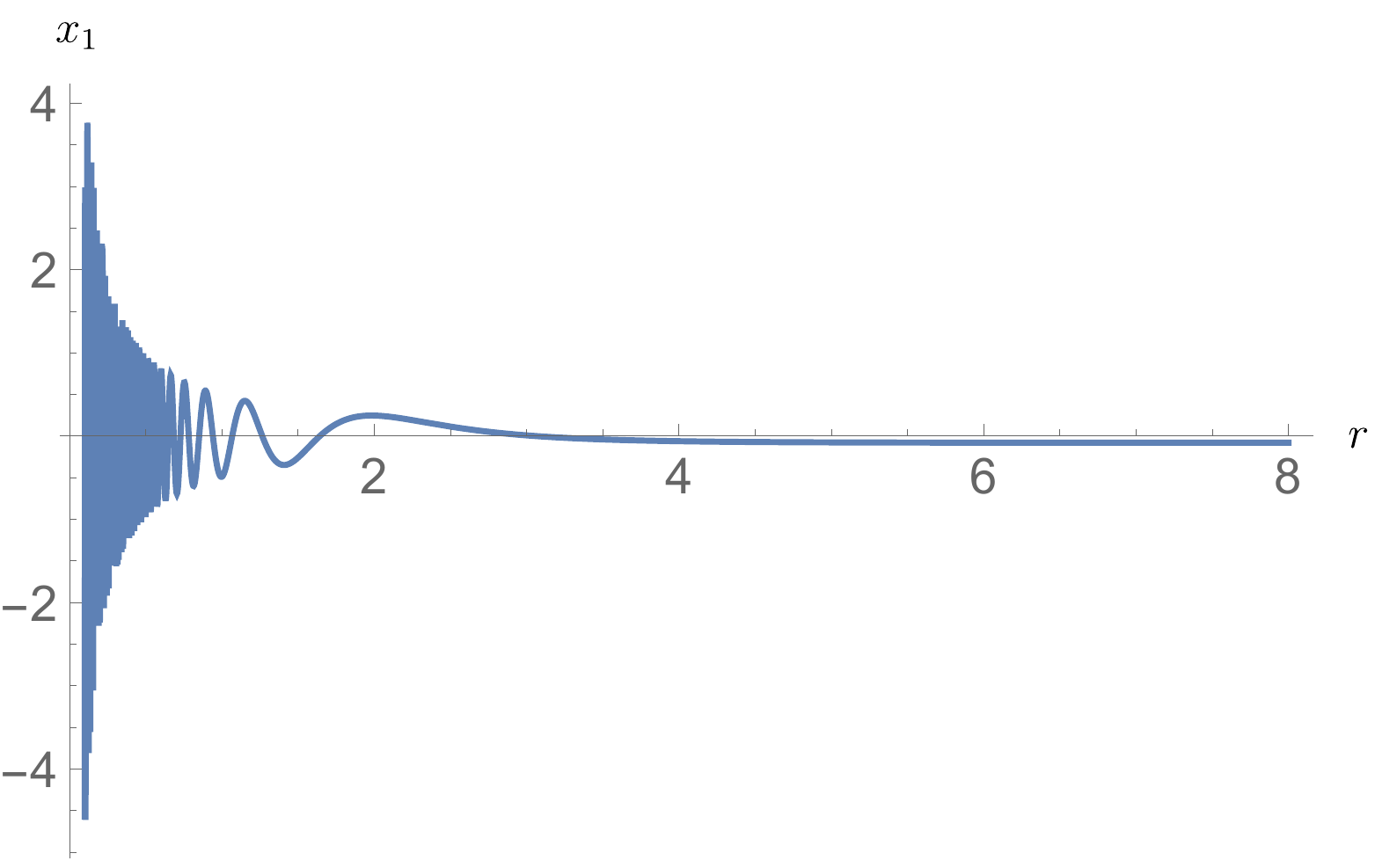}}
\caption{\small{The real part of the string world-sheet fluctuations in a fixed  Lifshitz background for a particular frequency and time. The large density of excitations in the deep IR $r$, get quickly smoothed out to result   the corresponding Brownian motion on the boundary  at large $r$.}}
\end{figure}

The two-point function of the fluctuations  in terms of the energy of the string in the low frequency limit
\bea\label{smallo}
\omega \ll \trr_b =\frac{r_b^{-\kappa}}{\kappa}~,
\eea
takes a very compact form
\bea\label{twopointfun01}
\vev{X_1(t) X_1(0)}~\sim~
E^\frac{4\kappa\prt{1-\nu}}{ \prt{a_0-\kappa}} ~|t|^{3-2\nu}~,\quad ~\rm{when}~\quad \nu \ge 1~,
\eea
and
\bea\label{twopointfun02}
\vev{X_1(t) X_1(0)}\sim E^0 ~|t|^{ 2\nu-1}~,\quad ~\rm{when}\quad~  \nu \le 1~,
\eea
where $X_1(t):=\delta x_1(t,r_b)$. The energy of the string comes by trading the cut-off $r_b$ at the boundary using the action of the static straight string
\bea\label{energy0}
E=\frac{1}{2\pi \a'}\int_{r_h}^{r_b} dr \sqrt{ -g_{00} g_{rr}}= \frac{1}{2 \pi\a'}  \prt{a_0-\kappa} r_b^{a_0-\kappa}~.
\eea
To summarise the analysis, we find that after an involved computation the two-point function \eq{twopointfun01}, \eq{twopointfun02} turns out to depend only on a single parameter, the order of the Bessel function. This is surprising and elegant result.  We  mention that in the unrelated studies of the chaos in the non-relativistic theories, it has been found that the order of the Bessel function of closed string solutions controls whether the theory is chaotic or not \cite{Giataganas:2014hma}. This is another example where the order of Bessel function plays such a significant role.

Depending on the properties of the dual field theory of \eq{polymetric1}, the two-point function of the particle fluctuations is characterized by one of the two branches we derived, where one of them is always independent of the mass of the quark. The null energy condition (NEC) is satisfied for regions in both branches of the two-point function  \eq{twopointfun01} and \eq{twopointfun02}, ensuring that both of them can be physical. For theories giving $\nu=3/2$ and $\nu=1/2$, where the AdS background belongs, we have a minimal rate of logarithmic growth, while for theories giving $\n=1$ we have the maximum rate of growth.

\subsubsection{Linear Response Function Analysis and Fluctuation-Dissipation Relation}

The response function of the system due to an applied external force $F(t)= E e^{-i\omega t} F(\omega)$ can be found from
\bea\label{response}
\vev{X_1(\omega)}=\chi(\omega) F(\omega)~,
\eea
where $\chi(\omega)$ is the admittance of the system. The force corresponds to a new boundary term
\bea\label{bboundf1}
 \frac{g_{11} \sqrt{-g_{00}}}{\sqrt{g_{uu}}}\delta x_1'(r_b)= 2\pi\alpha' F(t)~,
\eea
which does not modify the solution of string fluctuations, only its constants.   Making a tortoise coordinate transformation $r \rightarrow \tilde{r}$ we see that the fluctuation dynamics
in the  IR limit $r\rightarrow 0 \Rightarrow\tilde{r} \rightarrow \infty$ for $\kappa>0$, behaves like a wave equation in the flat space, where by choosing the ingoing boundary conditions we get
\bea\label{solamodes3}
\delta x_1(t,r)=e^{-i\omega t} g_\omega(r)~,\qquad  g_\omega(r)=r^{-\n\kappa}   H_\n\prt{\om \tilde{r}}~,
\eea
with $H:=H^{(1)}=J+iY$ being the Hankel function. Therefore the response function can be found to be
\bea\label{response00}
\chi(\omega) = \frac{2\pi\a'}{\omega} r_b^{-a_1} \frac{H_{\nu }(\omega \tilde{r}_b)}{H_{\n-1}(\omega \tilde{r}_b)}~.
\eea
All the information of the system and the theory are incorporated in the response function in an elegant way through the definition of $\n$ and $\trr$. The fluctuation-dissipation theorem can be also shown that it is satisfied
\bea\label{flds}
 \mbox{Im}\chi(\omega)=\frac{4 \alpha'}{ \omega^2}\frac{r_b^{-a_1}}{\tilde{r}_b}~\prt{J_{\nu-1}^2\prt{\omega \tilde{r}_b}-Y_{\nu-1}^2\prt{\omega \tilde{r}_b}}^{-1}=\frac{1}{2} \vev{X_1(\omega)X_1(0)}~,
\eea
for the theories fitting in our study.

By expanding the response function in the positive $\nu$ region
\bea
\chi(\omega)\sim -c_1\prtt{ m\prt{i\omega}^2 +\cO(\omega^4) +\gamma \prt{-i\omega}^{2\nu}+\cO(\omega^{2\nu+2})}^{-1}~,
\eea
we obtain the inertial mass $m$ and   the self-energy $\g$ of the particle
\bea\label{massin1}
m=\frac{r_b^{2\kappa\prt{\nu-1}}} {2 \kappa \prt{\nu-1}}~,~\qquad ~ \g=\frac{1-i\tan\prtt{\prt{\nu-\frac{1}{2}}\pi}}{\prt{ \prt{2 i \kappa}^{2\nu-1}\G\prt{\nu}^2}} \pi ~,
\eea
to find that the mass is not simply equal to the energy of the string given by \eq{energy0}. The order of the Bessel function plays a significant role in the result,   indicating for example which term dominates in the expression. For $\n>1$ the inertial mass dominates over the self energy at low frequency, otherwise the self energy is the dominant contribution.

\subsection{Thermal Diffusion of a Heavy Quark}

Let us now consider the Brownian motion in a thermal environment. The metric \eq{polymetric1} has to contain a black hole
\bea\label{polymetric2}
g_{00}=r^{a_0} f(r)~,\quad g_{rr}=\frac{1}{r^{a_u} f(r)}~,\quad g_{ii}=r^{a_i}~,\quad f(r):=1-\frac{r_h^{a_f}}{r^{a_f}}~,\quad r_h\neq 0~,\quad T=\frac{a_f}{4\pi}r_h^{\kappa}~,
\eea
where $a_f$ is a constant and $T$ is the temperature of the heat bath. The form of the blackening factor does not need to be necessarily chosen as the function above, for most of our analysis would be enough to assume a single zero at the position of the horizon $r_h$.

The equations of motion for the thermal fluctuations are generated by \eq{modes0} and can be brought to the Schr\"{o}dinger-like form by   a coordinate transformation presented in Appendix of  \cite{Giataganas:2018ekx} to get
\bea\label{schrodinger}
\frac{\partial^2 y}{\partial r_\star^2}+\prt{\omega^2-V(r)} y=0~,
\eea
where
\bea\label{rrstar}
&& y=h_\omega(r) r^\frac{a_1}{2}~ ,\qquad r_\star= -\frac{r_h^{-\kappa}}{a_f} ~B_{\prt{\frac{r_h}{r}}^{a_f}}\prtt{\frac{\kappa}{ a_f},0} ~,\\\label{yeq1}
&& V(r)=\frac{a_1}{2} r^{2\kappa} f(r)\prt{\left(a_0+a_u+\ff{a_1}{2}-1\right)f(r)+r f'(r) }~,
\eea
where $B_z\prtt{a,b}$ is the incomplete beta function. The monodromy patching procedure can be applied to obtain an approximate solution patching the regions between the black hole horizon all the way towards the boundary \cite{Motl:2003cd,Maldacena:1996ix,Harmark:2007jy,deBoer:2008gu,Tong:2012nf}. The three regions are chosen as
\bea
\label{rr1}
&&A)\qquad r\sim r_h\qquad\mbox{and}\qquad V(r)\ll \omega^2~ \Longleftrightarrow~ f(r)\ll \omega~.\\
\label{rr2}
&&B)\qquad V(r)\gg \omega^2~ \Longleftrightarrow~ f(r)\gg \omega~.\\
&&C)\qquad r\rightarrow \infty ~\Longleftrightarrow~ r\gg r_h~, \label{rr3}
\eea
giving the solutions for the Fourier modes $\delta x_1(t,r)=e^{-i\omega t} h_{\omega}(r)$ as \cite{Giataganas:2018ekx}
\bea\nn
&& h_{Ah}(r)=c_1\left(1- \frac{i\omega r_h^{-\kappa}}{a_f} \log\prt{\frac{r}{r_h}-1}\right)~,\\\label{monodromy1}
&& h_{Bh}(r)=c_3+c_4~ c_0+\frac{c_4}{a_f} r_h^{-2\kappa\n} \log\left(\frac{r}{r_h}-1\right)~,\\ \nn
&& h_{Bb}(r)=c_3-\frac{  c_4 }{2\kappa\n} r^{-2\kappa\n}~,\\ \nn
&& h_{Cb}(r)=c_5+c_6 ~r^{-2\kappa\nu}\left(\frac{\omega ~ \mbox{sign}\prt{1+2\kappa}}{2\kappa}\right)^{2\nu}~,
\eea
where the notation of the subscript used to identify which region of the space the solution covers: three regions and   the proximity close to (h)orizon or (b)oundary is labeled. By patching the solutions we eventually obtain for their constants
\bea\label{combineall}
c_5=c_1\prt{1+ i c_0 \om  r_h^{a_1} }~,\qquad c_6 = \frac{  i\om  r_h^{a_1}c_1}{2\kappa\n} \prt{ \frac{2\kappa}{\om ~ \mbox{sign}\prt{1+2\kappa}}}^{2\n}~,
\eea
to give the diffusive nature of the solution near the boundary
\bea \label{hcresult}
h_\omega(r)=c_1\left(1+i\omega c_0 r_h^{a_1}+\frac{i \omega r_h^{a_1}}{2\kappa\nu} r^{-2\kappa\nu}\right)~.
\eea

\subsubsection{Response Function Analysis, Fluctuation-Dissipation and Nature of the Thermal Noise}
The response function expansion for low frequency is given by \cite{Giataganas:2018ekx}
\bea
\chi(\omega)=2\pi \a'\prt{\frac{i}{\g \omega}-\frac{m}{\g^2}+\cO(\omega)}~,
\eea
where the damping coefficient and inertial mass reads
\bea\label{thermalcore1}
\g= r_h^{a_1}~,\qquad m= r_h^{2 a_1} \prt{-c_0+ \frac{r_b^{-2\kappa\n}}{2\kappa\n}}+ m_0~.
\eea
The mass receives a thermal correction compared with the zero temperature result $m_0$ \eq{massin1}. The diffusion constant in terms of the response function reads
\bea \label{diffusion1}
D=T \lim_{\om\rightarrow 0}\prt{-i~\om \chi(\om)}~,
\eea
and is found to depend on powers of temperature specified by the order of the Bessel function \cite{Giataganas:2018ekx}
\bea\label{ordiffusiont}
D=2\pi\a'  \prt{\frac{4\pi}{a_f}}^{2\n-1}~ T^{2\prt{1-\n}}~.
\eea
The monotonicity of the coefficient with respect to temperature, is increasing for $\nu<1$  and decreasing for $\n>1$, depending therefore on the characteristics of the dual field theory.

An interesting observation at the low frequency limit, is that the response function takes a universal form depending on the metric element along the spatial direction that the fluctuations occur and the temperature of the heat bath, given by the simple formula \cite{Giataganas:2018ekx}
\bea\label{responsemet1}
\chi(\om)=\frac{2\pi \a'}{-i\om g_{11}(r_h)}~.
\eea
The fluctuation-dissipation theorem is found to hold  when the density of states is $\sim \log\e/(4\pi^2 T)$. Moreover, it can be seen that the correlator of the random force $\xi$ is independent of the frequency $\omega$ indicating a white noise which depends on the temperature with powers of the Bessel function order $\vev{\xi~ \xi}\sim T^{2\nu}$.

\subsection{Dp-brane Fluctuations in a $d+1$-spacetime}

In \cite{Giataganas:2015ksa} it has been noticed that the dynamics of the $k$-strings corresponding holographically to certain Dp-branes, can be mapped to the dynamics of fundamental strings in theories that have a certain preserved quantities under a T-duality. Here we consider another type of special branes, the rigid ones, and study the  fluctuation and dissipation of the branes.

The brane is parametrized in the radial gauge as: $x_0=\tau,~r=\sigma_1:=\sigma,~ x_2=\sigma_2,\ldots,~ x_p=\sigma_p, ~ x_j=x_j(\tau,\sigma_i)$,  $j=1, p+1,\ldots,d$.  The fluctuations along the spatial transverse direction $x_1$ are given by the perturbation of the Dirac-Born-Infeld (DBI) action as \cite{Giataganas:2018ekx}
\bea
S_{\mbox{DBI},2}=\frac{T_p }{2} \int d\tau d\sigma_i \left(-\frac{g_{11} \sqrt{-g_{00}}\sqrt{g_{22}\ldots g_{p p}}}{\sqrt{g_{rr}}}\delta x_1'^2+ \frac{g_{11}\sqrt{g_{rr}}\sqrt{g_{22}\ldots g_{pp}}}{\sqrt{-g_{00}}}\delta \dx_1^2\right)~.
\eea
giving the mode equation for $\delta x_1 (\tau,\sigma)=e^{-i\omega \tau} h_\omega(r)$
\bea \label{bbmodes01}
\partial_r\prt{-\frac{g_{11} \sqrt{-g_{00}}\sqrt{g_{22}\ldots g_{pp}}}{\sqrt{g_{rr}}}h_\omega(r)'}- \frac{g_{11}\sqrt{g_{rr}}\sqrt{g_{22}\ldots g_{p p }}}{\sqrt{-g_{00}}}\omega^2 h_\omega(r)=0~.
\eea
We observe that under a  mapping of the form \cite{Giataganas:2018ekx}
\bea\label{mapping0}
g_{11}  \longrightarrow   g_{11}\sqrt{g_{22} g_{33}\ldots g_{dd} }~, ~~~\mbox{or equivalently } ~~a_1\longrightarrow \tilde{a}_1=a_1+\frac{1}{2}\prt{a_2+\ldots+a_p}~,
\eea
the string fluctuation equations \eq{modes0} become equivalent to the brane fluctuation equations. By defining the shifted constants
\bea\label{branedefinition}
\tilde{r}:=\frac{2 r^{\frac{1}{2}\left(2-\a_0-\a_u\right)}}{a_0+a_u-2}=\frac{r^{-\kappa}}{\kappa}~,\qquad \kappa:= \frac{a_0+a_u}{2}-1~,\qquad \tilde{\nu}:=\frac{a_0+2 \ta_1+a_u-2}{2\prt{a_0+a_u-2}}~,
\eea
the brane-related analysis follows in a straightforward way. For example the two-point function reads
\bea\label{branetwopointbf1}
\vev{X_1(t) X_1(0)}\sim
E^\frac{4\kappa\prt{1-\tilde{\nu}}}{ \left(a_0-\kappa\right)} ~|t|^{3-2\tilde{\nu}}~,\quad~ \rm{when}\quad \tilde{\nu} \ge 1~,
\eea
and
\bea\label{branetwopointbf2}
\vev{X_1(t) X_1(0)}\sim E^0 ~|t|^{ 2\tilde{\nu}-1}~,\quad \rm{when}\quad~ \tilde{ \nu } \le 1~.
\eea
The response function of the quantum brane fluctuations is
\bea\label{bform1}
\chi(\omega) =\frac{2\pi\alpha'}{\omega} r_b^{-a_1-\frac{1}{2}\prt{a_2+\ldots+a_p}} \frac{H_{\tilde{\nu} }(\omega \trr_b)}{H_{\tilde{\nu}-1}(\omega \tilde{r}_b)}~.
\eea
with a low-frequency expansion that provides the inertial mass $m$ and  the self-energy $\g$ of the state
\bea\label{braneform}
m= \frac{r_b^{2\kappa\left(\tilde{\nu}-1\right)}} {2 \kappa \prt{\tilde{\nu}-1}}~,\qquad ~\g= \frac{1-i\tan\prtt{\prt{\tilde{\nu}-\frac{1}{2}}\pi}}{\prt{ \prt{2 i \kappa}^{2\tilde{\nu}-1}\Gamma\left(\tilde{\nu}\right)^2}} \pi ~.
\eea
The analysis of the thermal fluctuations goes along the same lines with the string's one. For example for the case of branes the diffusion constant in terms of the temperature reads
\bea\label{bdiffusionbf}
D=2\pi\alpha'  \prt{\frac{4\pi}{a_f}}^{2\tilde{\nu}-1}~ T^{2\prt{1-\tilde{\nu}}}~,
\eea
where  for  $\tilde{\nu}<1$ increases with temperature and otherwise decreases.

For the branes in an arbitrary background the response function was proposed to take the form \cite{Giataganas:2018ekx}
\bea\label{bchibf}
\chi(\om)=\frac{2\pi \a'}{-i\om g_{11}(r_h)\sqrt{g_{22}(r_h)\ldots g_{pp}(r_h)}}~.
\eea

\subsection{Application of the Generic Formalism to Particular Theory}

Let us briefly demonstrate how the methodology described, applies to theories dual to the anisotropic black hole background found recently in Einstein-Axion-Dilaton action and contain as particular case the IIB supergravity background \cite{Giataganas:2017koz}. The geometry which   accommodates a black hole,  written in the form of \cite{Giataganas:2017koz,Giataganas:2018ekx} reads
\bea\label{hyscametr}
ds^2=a^2 C_R e^{\frac{\phi(r)}{2}} r^{-\frac{2\theta}{dz}} \left(r^{2} \prt{-f(r)dt^2+dx_i^2}+C_Z r^{\frac{2}{z}} dx_3^2+\frac{dr^2}{f(r)a^2 r^2}\right) ~,
\eea
where
\bea\label{fdilela}
 f(r)=1-\left(\frac{r_h}{r}\right)^{d+\prt{1-\theta}/z}~,\qquad e^{\frac{\phi(r)}{2}}=r^\frac{\sqrt{\theta^2+3 z\prt{1-\theta}-3}}{\sqrt{6}z}~
\eea
and
\bea
C_R=\frac{\prt{3z-\theta}\prt{1+3z-\theta}}{6 z^2}~,\qquad  C_Z=\frac{z^2}{2\prt{z-1}\prt{1+3z-\theta}}~.
\eea
The background becomes of IIB supergravity solution for $z=3/2,~\theta=0$ reproducing the geometry \cite{Azeyanagi:2009pr} and the IR geometry of the RG flow \cite{Mateos:2011ix}. The Hawking temperature of the theory can be found equal to
\bea
T=\frac{|d+(1-\theta)/z|}{4 \pi r_h^z} ~.
\eea
Let us work with $d=3$ spatial dimensions. The first task is to determine the order of the Bessel functions for fluctuations along each direction by using \eq{definitionn1}
\bea
&&\nu_1=\frac{18z-4\theta +\sqrt{6}\sqrt{3z\prt{1-\theta}-3+\theta^2}}{12 z}~ ,\\
&&\nu_3=\frac{12+6z-4\theta +\sqrt{6}\sqrt{3z\prt{1-\theta}-3+\theta^2}}{12 z}~.
\eea
The fluctuations along the $x_1$ and $x_3$ behave in a different manner. The two-point function along $x_1$ is
\bea\label{twopointa1}
\vev{X_1(t) X_1(0)}\sim E^{-2} ~|t|^{3-2\nu_1}~,\quad \rm{when}\quad \nu_1 \ge 1  ~,
\eea
where for $\n_1<1$ there is no physical and stable theory. The $x_3$ fluctuations give
\bea\label{twopointa4}
\vev{X_3(t) X_3(0)}\sim
E^{2\frac{12-6z-4\theta +\sqrt{6}\sqrt{3z\prt{1-\theta}-3+\theta^2}}{-6z+4\theta-\sqrt{6}\sqrt{3z\prt{1-\theta}-3+\theta^2}}} ~|t|^{3-2\nu_3}~,\quad \rm{when}\quad \nu_3 \ge 1 ~,
\eea
or
\bea
\vev{X_3(t) X_3(0)}\sim E^0 ~|t|^{ 2\nu_3-1}~,\quad \rm{when}\quad  \nu_3 \le 1~.
\eea
The diffusion constant depends on the direction and is given by
\bea
D_i=2\pi \alpha'\prt{\frac{4\pi}{d+\prt{1\-\theta}/z}}^\prt{2\nu_i-1}T^{2\prt{1-\nu_i}}~,
\eea
where $i=1,3$ labels the direction of the fluctuations.

\section{Stochastic Motion of the Moving Heavy Quark: Theory Independent Approach}

Let us consider a heavy particle moving with a velocity $v$ in a strongly coupled environment. The Langevin coefficients $\kappa_\perp, \kappa_\parallel$, corresponding to  the mean squared momentum per unit of time in transverse and parallel direction with respect to the quark's motion, are obtained by analyzing the fluctuation of the trailing Wilson line. The out of equilibrium relativistic heavy quarks go under a Brownian motion
with a stochastic force $\xi(t)$. The  methodology for generic theories including the
anisotropic ones , was developed in \cite{Giataganas:2013hwa,Giataganas:2013zaa}  by providing a set of readily applicable formulas for the observables and extracting certain behaviors of them in a wide class of theories. The set of theories accommodated in this study include for example: conformal, non-relativistic, hyperscaling violation, anisotropic  and theories under certain magnetic fields.  The formulas of  \cite{Giataganas:2013hwa,Giataganas:2013zaa}  have been applied on several particular models,
for example in \cite{Finazzo:2016mhm}. The initiating works on the subject include \cite{Herzog:2006gh,Gubser:2006bz,Gubser:2006nz,CasalderreySolana:2006rq,CasalderreySolana:2007qw} and
early works that contributed to further development include \cite{Giecold:2009cg,Gursoy:2009kk,HoyosBadajoz:2009pv,Gursoy:2010aa}.

It is worthy to mention, that by studying the relativistic heavy quark diffusion in theories with rotational invariance  it  has been  found that a universal inequality for the Langevin coefficients exist
$\kappa_\parallel\ge \kappa_\perp$  \cite{Giataganas:2013hwa,Gursoy:2010aa}. Namely the longitudinal Langevin diffusion coefficient along the quark motion is larger compared to that of the transverse direction. This inequality has been proved to be violated in the presence of strongly coupled anisotropies \cite{Giataganas:2013hwa,Giataganas:2013zaa}, in a similar way with the well known shear viscosity over entropy density bound \cite{Kovtun:2004de,Rebhan:2011vd,Jain:2015txa,Giataganas:2017koz} (other relevant works include \cite{Erdmenger:2010xm,Samanta:2016pic,Ge:2015owa,Kolekar:2016pnr}).

Other related works on the drag and stochastic sting motion include  \cite{Gubser:2006qh,CasalderreySolana:2009rm,Giecold:2009cg,Rajagopal:2015roa,Herzog:2007kh,Akamatsu:2015kaa,Horowitz:2015dta,Nakamura:2013yqa, Arean:2016het,Roy:2009sw,Horowitz:2009pw,Caceres:2006as,Ahmadvand:2015gfi,Zhang:2018mqt} while dragging of the particle even at zero temperature in non-relativistic theories has been observed \cite{Hartnoll:2009ns,Kiritsis:2012ta,Fadafan:2009an}.  An earlier review with references therein is \cite{CasalderreySolana:2011us}.

\subsection{The Trailing String} \label{section:trailing}

Let us review first the theory-independent analysis of the trailing string following for the drag force analysis of the Appendices \cite{Giataganas:2012zy,Giataganas:2013lga} while for the stochastic motion analysis we follow \cite{Giataganas:2013hwa,Giataganas:2013zaa}.  We use a similar metric to \eq{gen1} with $u=1/r$ and therefore a boundary at $u\rightarrow 0$ and a black hole horizon at $u=u_h$
\bea\label{gen22}
ds^2=-g_{00}(u) dx_0^2+ g_{uu}(u)du^2 + \sum_{i=1}^{d} g_{ii}(u) dx_i^2 ~.
\eea
The string parametrization for a quark moving on the $x_1$ direction with velocity $v$, is  $x_0=\tau,~~u=\sigma,~~x_1=v~t+\xi(u)$, where $\xi$ is the profile of the string in the bulk, satisfying on the boundary $\xi(u_b)=0$.
The equation of motion is
\bea\label{trailings}
\xi'^2=-g_{uu} C^2\,\frac{g_{00}+g_{11}\,v^2} {g_{00}g_{11}\prt{C^{2}+g_{00}g_{11}}}~,~\qquad ~ C:=2~\pi\,\alpha'\,\Pi^1_u~,
\eea
where $u_0$ is the black hole horizon of the world-sheet metric given by the solution of
\bea\label{wshorizon}
g_{00}(u_0)=-g_{11}(u_0)\,v^2~,
\eea
when $g_{uu}(u_0)\neq 0$. For $v=0$ the horizon of the worldsheet coincides with the horizon of the black hole, satisfying the natural expectations. The dragging of the particle is given by \cite{Giataganas:2012zy}
\bea\label{drag1}
F_{drag,x_1}=-\frac{1}{2\pi\alpha'}\frac{\sqrt{-g_{00}(u_0)\,g_{11}(u_0)}} {2\pi}=-\frac{ g_{11}(u_0)~v}{2\pi\alpha'}~,
\eea
while the friction coefficient is defined by
\bea\label{drag22}
F_{drag}=\frac{dp}{dt}=-\eta_D p~,\qquad \eta_D=\frac{g_{11}(u_0)}{2\pi\alpha' M \gamma}~,
\eea
where $M$ is the mass of the heavy quark, $p=M\,v~\gamma$ and $\g$ is the Lorentz factor.
The worldsheet has a blackening factor, and therefore a temperature is associated to it. To find the temperature we diagonalize the world-sheet metric to get \cite{Giataganas:2013hwa}
\bea\label{tws1}
T_{ws}^2=
\frac{1}{16\pi^2}\bigg|\frac{1}{g_{00} g_{uu}}\prt{g_{00}~g_{11}}' \prt{\frac{g_{00}}{g_{11}}}'\bigg|\Bigg|_{u=u_0}
~.
\eea
$T_{ws}$ should be thought as the effective temperature that the quark feels and is the temperature that appears in the Einstein equations relating the diffusion and the Langevin coefficients. The effective temperature in most theories turns out to be lower than the heat bath temperature, although in anisotropic theories may become higher. The natural expectation for the static quark $(v=0)$ would be that it feels the heat bath temperature and this can be verified by the above relation to obtain $T_{ws}=T$.

\subsection{Fluctuation of the Moving Trailing String}

Let us review the fluctuations around the trailing string in a generic background. The method was developed in  \cite{Giataganas:2013hwa,Giataganas:2013zaa} and the action was found to be
\bea \label{actiontr22}
S_2=-\frac{1}{2\pi\alpha'}\int d\tau d\sigma \,\frac{H^{\alpha\beta}}{2}~\left[N(u)\,\pp_\alpha \delta x_1\,\partial_\beta \delta x_1+\sum_{i\neq 1}{g_{ii}}\partial_\alpha \delta x_i~\partial_\beta \delta x_i\right]~,
\eea
where $H^{\a\beta}=\sqrt{-h}{h}^{\alpha\beta},$ and  $h^{\alpha\beta}$ is inverse of the diagonalized induced world-sheet metric given by
\bea
h_{\sigma\sigma}
=\frac{g_{00}g_{uu}g_{11}}{g_{00}g_{11}+C^2}~,\qquad
h_{{\tau}{\tau}}=g_{00}+v^2\,g_{11}~.
\eea
Taking advantage of the membrane paradigm it has been found that the Langevin coefficients are computed by \cite{Giataganas:2013hwa,Giataganas:2013zaa}
\bea\label{membranek}
\kappa_\perp=\frac{1}{\pi\alpha'}\,g_{kk}\bigg|_{u=u_0} T_{ws}~,~\qquad~
\kappa_\parallel=\frac{1}{\pi\alpha'}\,\frac{\left(g_{00}g_{11}\right)'} {g_{11}\,\left(\frac{g_{00}}{g_{11}}\right)'}\Bigg|_{u=u_0} T_{ws}~,
\eea
where the index $k$ denotes a particular transverse direction to that of motion $x_1$ and no summation is taken. The ratio takes the surprisingly  compact form
\bea\label{ratioklkt}
\frac{\kappa_\parallel}{\kappa_\perp}=\frac{\left(g_{00}g_{11}\right)'} {g_{kk}g_{11}\,\left(\frac{g_{00}}{g_{11}}\right)'}\Bigg|_{u=u_0}~.
\eea
In isotropic spaces it has been found that $\kappa_\parallel>\kappa_\perp$ for any velocity of the quark's motion  \cite{Giataganas:2013hwa,Gursoy:2010aa}. In anisotropic theories the universal condition is violated \cite{Giataganas:2013hwa,Giataganas:2013zaa}. It exists a critical quark velocity  $v_c$ beyond which the inequality gets inverted to $\kappa_\parallel<\kappa_\perp$.

\subsection{Excitation of the Medium due to Heavy Quark Motion}

The quark deposits energy in the medium through its interactions with the environment, and as a result excitations in the medium occur  which are expected to be well described by linearized hydrodynamics. Due to the motion of the particle  through the plasma, a laminar wake is generated behind it, which has been shown to be of universal strength with respect to the total drag force exerted by the plasma. A sonic boom has been discovered for probes moving faster than the speed of sound and a diffusion wake behind the quark's motion has been also found \cite{Gubser:2007ga,Gubser:2007ni,CasalderreySolana:2006sq,Friess:2006fk}.

The total action to compute such backreacted effects is given by
\bea
S=\frac{1}{2 \kappa^2}\int d^5 x \sqrt{-g} \prt{R+\frac{12}{L^2}}-S_{NG}~,
\eea
where the NG is computed for the trailing string of the section \ref{section:trailing}. The equations of motion read
\bea
R_{\mu\nu}-\frac{1}{2} g_{\mu\nu} R-\frac{6}{L^2} g_{\mu\nu}=\tau_{\mu\nu}~.
\eea
If we concentrate on the AdS spacetime
\bea
ds^2=\frac{1}{u^2}\prt{-f(u)dt^2+d\vec{x}^2+\frac{du^2}{f(u)}}~,\qquad~ f(u)=1-\frac{u^4}{u_h^4}~,
\eea
the bulk stress-energy tensor for the trailing string takes the form
\bea
\tau_{\mu\nu}=-\frac{\kappa^2}{2\pi\alpha'}  u^3  \sqrt{1-v^2} \partial_\alpha x^\mu \partial^a x^\nu~,
\eea
computed on the string world-sheet. The backreaction of the string metric can be found by considering the fluctuations  $g_{\mu\nu}=g_{\mu\nu}^{0}+h_{\mu\nu}$ on the $g_{\mu\nu}^{0}$ AdS metric. The $h_{\mu\nu}$ depends for the trailing string in terms of $x_1-v t$ and we have the freedom to take the axial gauge $h_{\mu u}=0$. The system of equations consists of ten second order differential equations in $u$, minus five first order constraints, therefore we need to specify fifteen integration constants. These are fixed by imposing conditions on the boundary of the space and at the horizon of the black hole as in \cite{Gubser:2009sn}.

The low momentum asymptotics can be obtained analytically by formally expanding all variables in low momenta, where the diffusion pole and the sound pole expected from the hydrodynamic behavior of the plasma are confirmed by the computations \cite{Gubser:2009sn}. On the other hand at the large momenta, the leading term of the stress-energy tensor is a boosted version of the stationary quark's  as expected. At scales much shorter that the typical length scale of the fluid the quark does not see the plasma and feels a vacuum state. The next order reveals the existence of a critical velocity for the quark's motion that passes from a region of energy depletion behind the quark, to a region of energy depletion in front of it. The full numerical analysis is presented in \cite{Gubser:2009sn} for the various quark velocities.

\subsection{Non-Perturbative Monte Carlo Simulations of the Heavy Quark   Diffusion}

An estimate of the heavy quark momentum diffusion with Monte-Carlo simulation using the Backus-Gilbert method  \cite{bgflattice} was done in  \cite{Francis:2015daa} giving
\bea
\kappa=1.8 \ldots 3.4~ T^3~.
\eea
The result seems to be in agreement with a next to leading order computation in perturbative QCD using the hard thermal loop effective theory and by setting the scales and the coupling to the usual ones used in the heavy ion collisions \cite{CaronHuot:2007gq}. Then the diffusion coefficient is estimated to
\bea
D T= 0.35 \ldots 1.1
\eea
and has higher values compared to the one predicted for the light quarks in the continuum limit \cite{Ding:2010ga,Burnier:2012ts}. This can be naturally explained with the fact that the heavy quarks feel slightly weaker interactions. Using these methods, it would be interesting to show that the heavy mass limit is justified for the lighter charm quarks, and not only for the heavier bottoms quarks. A promising direction would also be to estimate the effects from dynamical quarks on the heavy quark diffusion, where the screening should affect the observables. To this direction the gauge/gravity duality could also provide very insightful qualitative results. In this context, the flavor is added to the correspondence with the use of the Dp-branes, where the quenched limit is  easier tractable \cite{Karch:2002sh,Erdmenger:2007cm}, while the unquenched limit is demanding computationally \cite{Burrington:2004id,Nunez:2010sf}, so approximate or numerical methods have been developed and the relevant screening on several observables has been observed, for example \cite{Kirsch:2005uy,Giataganas:2011nz,Bigazzi:2014qsa,Alho:2012mh,Faedo:2017aoe,Li:2016gtz}.

\section{Summary}

In this brief review we have presented a theory independent treatment of stochastic heavy quark dynamics. In the introduction we have justified why the heavy quarks admit a stochastic treatment. Then we have presented the analysis of quantum and thermal fluctuations for a static quark by considering fluctuations of the straight string. We have moved on to the analysis of the trailing string fluctuations to obtain the Langevin equations. The idea of this review is to present the model independent holographic formulas applicable to wide class of theories, which have been obtained in \cite{Giataganas:2018ekx} for the static quark and in \cite{Giataganas:2013hwa,Giataganas:2013zaa} for the trailing string. We briefly summarize most of the formulas below.

\textbf{Quantum Fluctuations of the Heavy Particle:} Using the wide class of theories described by \eq{gen1} we obtain the generic form of the action \eq{oactionorder2} describing the fluctuations with the mode equation \eq{modes0}. For the theories of the form \eq{polymetric1}  the fluctuations are given  by a Bessel type solution  \eq{solutionmodes1} with order $\nu$ \eq{definitionn1}
which depends on the background geometry. By applying the boundary conditions and appropriate quantization we determine the constants of the solutions. The two point-function of the fluctuation has a surprisingly compact form and the two branches  \eq{twopointfun01} and \eq{twopointfun02}, controlled by the value of the Bessel function and therefore by the properties of the theory we study.

The response function analysis is done by applying the generic boundary force  \eq{bboundf1}. The modification of the boundary condition leads to a different solution for the fluctuations specified by \eq{solamodes3}. The response function is found in terms of the Hankel function with the same order $\nu$. Then the fluctuation-dissipation theorem is found to be satisfied \eq{flds}. By expanding the response function we determine the inertial mass and the self energy of the particle for the whole class of theories \eq{massin1}. All the results and their properties heavily depend on the order $\n$ incorporating the information of the theory.

\textbf{Thermal Diffusion of the Heavy Particle:} Including to our geometry the black hole \eq{polymetric2}, we study the thermal string fluctuations. The solution to the equation of fluctuations \eq{schrodinger} is  involved and the monodromy patching method needs to be used, patching certain approximate solutions in different regions along the holographic direction \eq{monodromy1}. The solution close to the boundary is given by \eq{hcresult} depending heavily of the asymptotics of the metric element along the fluctuations and the Bessel function order. The response function is found to take the form \eq{responsemet1}, exclusively depending on the black hole horizon along the direction of fluctuations. The self energy and the thermally corrected inertial mass are given by \eq{thermalcore1}. Interestingly the diffusion coefficient scales with the temperature in way that is solely controlled by the Bessel function order \eq{ordiffusiont}, realizing how elegantly the information of the wide class backgrounds is encoded in the order $\nu$.

\textbf{Diffusion of a Rigid type Dp-Branes:} By mapping the equations of the rigid branes \eq{bbmodes01} to the string fluctuations we find a way to read all the stochastic observables and coefficients from the analysis done on the strings. The mapping of the string to brane fluctuation is implemented by \eq{mapping0} giving the shifted Bessel function order \eq{branedefinition}. Then the two-point function of Dp-branes \eq{branetwopointbf1}, \eq{branetwopointbf2}, the response function \eq{bchibf}, the inertial mass and self energy \eq{braneform}, and the diffusion coefficient \eq{bdiffusionbf} can be read in a straightforward way from the prescription explained above.

\textbf{Dragging of the Heavy Moving Particle:} We consider a heavy particle moving with a velocity $v$ in a strongly coupled environment described by a metric with elements being arbitrary functions of the holographic direction  \eq{gen22}. The profile of the trailing string is given by the equation \eq{trailings} and the two-dimensional worldsheet has a black hole with a horizon given by the solution of the equation \eq{wshorizon}. The drag force is expressed in terms of the metric element of the direction of the quark's motion \eq{drag1}, while the friction coefficient is found in \eq{drag22}. The moving quark feels an effective temperature different than the one of the heat bath given by the equation \eq{tws1}.

\textbf{Langevin Coefficients of the Heavy Moving Particle:} The action of the fluctuations is given by \eq{actiontr22}. Employing the membrane paradigm we obtain the Langevin coefficients given in terms of the metric elements and the effective temperature that the quark feels \eq{membranek}. These are generic powerful formulas directly applicable to the thermal holographic theories. The ratio of the two coefficients take a very simple form given by \eq{ratioklkt} and turns out that for isotropic theories it satisfies a universal relation inequality of the form $\kappa_\parallel>\kappa_\perp$ for the whole range of the quark motion velocity. In anisotropic theories the universal inequality gets inverted for a critical speed of the quark, and the universality is violated in a similar  way that the viscosity over entropy bound is violated in anisotropic theories.

~\newline~\newline
\textbf{Acknowledgements:} The work of the author is supported by the National Center of Theoretical Science (NCTS) and the grants 101-2112-M-007-021-MY3 and 104-2112-M-007 -001 -MY3 of the Ministry of Science and Technology of Taiwan (MOST).  This review submitted for the Proceedings of the Corfu Summer Institute 2017 'School and Workshops on Elementary Particle Physics and Gravity', 2-28 September 2017, Corfu, Greece.

\bibliographystyle{JHEP}

\begin{thebibliography}{100}

\bibitem{brown01}
R.~Brown and J.~J. Bennett, \emph{The miscellaneous botanical works of Robert
  Brown}, vol.~1. London,Published for the Ray society by R. Hardwicke,.

\bibitem{ein1905}
A.~{Einstein}, \emph{{{\"U}ber die von der molekularkinetischen Theorie der
  W{\"a}rme geforderte Bewegung von in ruhenden Fl{\"u}ssigkeiten suspendierten
  Teilchen}}, \href{https://doi.org/10.1002/andp.19053220806}{\emph{Annalen der
  Physik} {\bfseries 322} (1905) 549}.

\bibitem{sutherland1}
W.~Sutherland, \emph{Lxxv. a dynamical theory of diffusion for non-electrolytes
  and the molecular mass of albumin},
  \href{https://doi.org/10.1080/14786440509463331}{\emph{The London, Edinburgh,
  and Dublin Philosophical Magazine and Journal of Science} {\bfseries 9}
  (1905) 781}
  [\href{https://arxiv.org/abs/https://doi.org/10.1080/14786440509463331}{{\ttfamily
  https://doi.org/10.1080/14786440509463331}}].

\bibitem{smol1}
von Smoluchowski~M., \emph{Zur kinetischen theorie der brownschen
  molekularbewegung und der suspensionen},
  \href{https://doi.org/10.1002/andp.19063261405}{\emph{Annalen der Physik}
  {\bfseries 326} 756}
  [\href{https://arxiv.org/abs/https://onlinelibrary.wiley.com/doi/pdf/10.1002/andp.19063261405}{{\ttfamily
  https://onlinelibrary.wiley.com/doi/pdf/10.1002/andp.19063261405}}].

\bibitem{perrin1}
J.~Perrin, \emph{{Mouvement brownien et r\'{e}alit\'{e} mol\'{e}culaire}},
  {\emph{Annales de Chimie et de Physique} {\bfseries 18} (1909) 5}.

\bibitem{brownianII}
M.~C. Wang and G.~E. Uhlenbeck, \emph{On the theory of the brownian motion ii},
  \href{https://doi.org/10.1103/RevModPhys.17.323}{\emph{Rev. Mod. Phys.}
  {\bfseries 17} (1945) 323}.

\bibitem{brownian100}
P.~Hanggi and F.~Marchesoni, \emph{Introduction: 100years of brownian motion},
  \href{https://doi.org/10.1063/1.1895505}{\emph{Chaos: An Interdisciplinary
  Journal of Nonlinear Science} {\bfseries 15} (2005) 026101}
  [\href{https://arxiv.org/abs/https://doi.org/10.1063/1.1895505}{{\ttfamily
  https://doi.org/10.1063/1.1895505}}].

\bibitem{brownianrel}
J.~Dunkel and P.~Hanggi, \emph{Relativistic brownian motion},
  \href{https://doi.org/https://doi.org/10.1016/j.physrep.2008.12.001}{\emph{Physics
  Reports} {\bfseries 471} (2009) 1 }.

\bibitem{brownian111}
X.~Bian, C.~Kim and G.~E. Karniadakis, \emph{111 years of brownian motion},
  \href{https://doi.org/10.1039/C6SM01153E}{\emph{Soft Matter} {\bfseries 12}
  (2016) 6331}.

\bibitem{Svetitsky:1987gq}
B.~Svetitsky, \emph{{Diffusion of charmed quarks in the quark-gluon plasma}},
  \href{https://doi.org/10.1103/PhysRevD.37.2484}{\emph{Phys. Rev.} {\bfseries
  D37} (1988) 2484}.

\bibitem{vanHees:2004gq}
H.~van Hees and R.~Rapp, \emph{{Thermalization of heavy quarks in the
  quark-gluon plasma}},
  \href{https://doi.org/10.1103/PhysRevC.71.034907}{\emph{Phys. Rev.}
  {\bfseries C71} (2005) 034907}
  [\href{https://arxiv.org/abs/nucl-th/0412015}{{\ttfamily nucl-th/0412015}}].

\bibitem{Moore:2004tg}
G.~D. Moore and D.~Teaney, \emph{{How much do heavy quarks thermalize in a
  heavy ion collision?}},
  \href{https://doi.org/10.1103/PhysRevC.71.064904}{\emph{Phys. Rev.}
  {\bfseries C71} (2005) 064904}
  [\href{https://arxiv.org/abs/hep-ph/0412346}{{\ttfamily hep-ph/0412346}}].

\bibitem{Mustafa:2004dr}
M.~G. Mustafa, \emph{{Energy loss of charm quarks in the quark-gluon plasma:
  Collisional versus radiative}},
  \href{https://doi.org/10.1103/PhysRevC.72.014905}{\emph{Phys. Rev.}
  {\bfseries C72} (2005) 014905}
  [\href{https://arxiv.org/abs/hep-ph/0412402}{{\ttfamily hep-ph/0412402}}].

\bibitem{Giataganas:2018ekx}
D.~Giataganas, D.-S. Lee and C.-P. Yeh, \emph{{Quantum Fluctuation and
  Dissipation in Holographic Theories: A Unifying Study Scheme}},
  \href{https://arxiv.org/abs/1802.04983}{{\ttfamily 1802.04983}}.

\bibitem{deBoer:2008gu}
J.~de~Boer, V.~E. Hubeny, M.~Rangamani and M.~Shigemori, \emph{{Brownian motion
  in AdS/CFT}},
  \href{https://doi.org/10.1088/1126-6708/2009/07/094}{\emph{JHEP} {\bfseries
  0907} (2009) 094} [\href{https://arxiv.org/abs/0812.5112}{{\ttfamily
  0812.5112}}].

\bibitem{Son:2009vu}
D.~T. Son and D.~Teaney, \emph{{Thermal Noise and Stochastic Strings in
  AdS/CFT}}, \href{https://doi.org/10.1088/1126-6708/2009/07/021}{\emph{JHEP}
  {\bfseries 07} (2009) 021} [\href{https://arxiv.org/abs/0901.2338}{{\ttfamily
  0901.2338}}].

\bibitem{CaronHuot:2011dr}
S.~Caron-Huot, P.~M. Chesler and D.~Teaney, \emph{{Fluctuation, dissipation,
  and thermalization in non-equilibrium AdS5 black hole geometries}},
  \href{https://doi.org/10.1103/PhysRevD.84.026012}{\emph{Phys. Rev.}
  {\bfseries D84} (2011) 026012}
  [\href{https://arxiv.org/abs/1102.1073}{{\ttfamily 1102.1073}}].

\bibitem{Sonner:2012if}
J.~Sonner and A.~G. Green, \emph{{Hawking Radiation and Non-equilibrium Quantum
  Critical Current Noise}},
  \href{https://doi.org/10.1103/PhysRevLett.109.091601}{\emph{Phys. Rev. Lett.}
  {\bfseries 109} (2012) 091601}
  [\href{https://arxiv.org/abs/1203.4908}{{\ttfamily 1203.4908}}].

\bibitem{Tong:2012nf}
D.~Tong and K.~Wong, \emph{{Fluctuation and Dissipation at a Quantum Critical
  Point}}, \href{https://doi.org/10.1103/PhysRevLett.110.061602}{\emph{Phys.
  Rev. Lett.} {\bfseries 110} (2013) 061602}
  [\href{https://arxiv.org/abs/1210.1580}{{\ttfamily 1210.1580}}].

\bibitem{Hubeny:2010ry}
V.~E. Hubeny and M.~Rangamani, \emph{{A Holographic view on physics out of
  equilibrium}}, \href{https://doi.org/10.1155/2010/297916}{\emph{Adv. High
  Energy Phys.} {\bfseries 2010} (2010) 297916}
  [\href{https://arxiv.org/abs/1006.3675}{{\ttfamily 1006.3675}}].

\bibitem{Fischler:2014tka}
W.~Fischler, P.~H. Nguyen, J.~F. Pedraza and W.~Tangarife, \emph{{Fluctuation
  and dissipation in de Sitter space}},
  \href{https://doi.org/10.1007/JHEP08(2014)028}{\emph{JHEP} {\bfseries 08}
  (2014) 028} [\href{https://arxiv.org/abs/1404.0347}{{\ttfamily 1404.0347}}].

\bibitem{Yeh:2014mfa}
C.-P. Yeh, J.-T. Hsiang and D.-S. Lee, \emph{{Holographic influence functional
  and its application to decoherence induced by quantum critical theories}},
  \href{https://doi.org/10.1103/PhysRevD.91.046009}{\emph{Phys. Rev.}
  {\bfseries D91} (2015) 046009}
  [\href{https://arxiv.org/abs/1410.7111}{{\ttfamily 1410.7111}}].

\bibitem{Yeh:2015cra}
C.-P. Yeh and D.-S. Lee, \emph{{Subvacuum effects in quantum critical theories
  from a holographic approach}},
  \href{https://doi.org/10.1103/PhysRevD.93.126006}{\emph{Phys. Rev.}
  {\bfseries D93} (2016) 126006}
  [\href{https://arxiv.org/abs/1510.05778}{{\ttfamily 1510.05778}}].

\bibitem{Banerjee:2015vmo}
P.~Banerjee, \emph{{Holographic Brownian motion at finite density}},
  \href{https://doi.org/10.1103/PhysRevD.94.126008}{\emph{Phys. Rev.}
  {\bfseries D94} (2016) 126008}
  [\href{https://arxiv.org/abs/1512.05853}{{\ttfamily 1512.05853}}].

\bibitem{Moerman:2016wpv}
R.~W. Moerman and W.~A. Horowitz, \emph{{A semi-classical recipe for wobbly
  limp noodles in partonic soup}},
  \href{https://arxiv.org/abs/1605.09285}{{\ttfamily 1605.09285}}.

\bibitem{Lee:2016wcn}
D.-S. Lee and C.-P. Yeh, \emph{{A holographic description of negative energy
  states}}, \href{https://doi.org/10.1007/JHEP09(2016)059}{\emph{JHEP}
  {\bfseries 09} (2016) 059}
  [\href{https://arxiv.org/abs/1606.02420}{{\ttfamily 1606.02420}}].

\bibitem{Kiritsis:2013iba}
E.~Kiritsis, L.~Mazzanti and F.~Nitti, \emph{{The confining trailing string}},
  \href{https://doi.org/10.1007/JHEP02(2014)081}{\emph{JHEP} {\bfseries 02}
  (2014) 081} [\href{https://arxiv.org/abs/1311.2611}{{\ttfamily 1311.2611}}].

\bibitem{Ho:2013rra}
S.-H. Ho, W.~Li, F.-L. Lin and B.~Ning, \emph{{Quantum Decoherence with
  Holography}}, \href{https://doi.org/10.1007/JHEP01(2014)170}{\emph{JHEP}
  {\bfseries 01} (2014) 170} [\href{https://arxiv.org/abs/1309.5855}{{\ttfamily
  1309.5855}}].

\bibitem{Fischler:2012ff}
W.~Fischler, J.~F. Pedraza and W.~Tangarife~Garcia, \emph{{Holographic Brownian
  Motion in Magnetic Environments}},
  \href{https://doi.org/10.1007/JHEP12(2012)002}{\emph{JHEP} {\bfseries 12}
  (2012) 002} [\href{https://arxiv.org/abs/1209.1044}{{\ttfamily 1209.1044}}].

\bibitem{Roychowdhury:2015mta}
D.~Roychowdhury, \emph{{Quantum fluctuations and thermal dissipation in higher
  derivative gravity}},
  \href{https://doi.org/10.1016/j.nuclphysb.2015.06.013}{\emph{Nucl. Phys.}
  {\bfseries B897} (2015) 678}
  [\href{https://arxiv.org/abs/1506.04548}{{\ttfamily 1506.04548}}].

\bibitem{landausp}
L.~D. Landau and E.~M. Lifshitz, \emph{{ Statistical Physics, Part I}},
  {\emph{Pergamon Press, Oxford} {\bfseries {Course of Theoretical Physics,
  Vol. 5, Third Edition }} (1980) }.

\bibitem{caldeira1983}
A.~Caldeira and A.~Leggett, \emph{Path integral approach to quantum brownian
  motion},
  \href{https://doi.org/https://doi.org/10.1016/0378-4371(83)90013-4}{\emph{Physica
  A: Statistical Mechanics and its Applications} {\bfseries 121} (1983) 587 }.

\bibitem{Schwinger}
J.~Schwinger, \emph{Brownian motion of a quantum oscillator},
  \href{https://doi.org/10.1063/1.1703727}{\emph{Journal of Mathematical
  Physics} {\bfseries 2} (1961) 407}
  [\href{https://arxiv.org/abs/https://doi.org/10.1063/1.1703727}{{\ttfamily
  https://doi.org/10.1063/1.1703727}}].

\bibitem{Feynman:1963fq}
R.~P. Feynman and F.~L. Vernon, Jr., \emph{{The Theory of a general quantum
  system interacting with a linear dissipative system}},
  \href{https://doi.org/10.1016/0003-4916(63)90068-X}{\emph{Annals Phys.}
  {\bfseries 24} (1963) 118}.

\bibitem{Grabert:1988yt}
H.~Grabert, P.~Schramm and G.~L. Ingold, \emph{{Quantum Brownian motion: The
  Functional inegral approach}},
  \href{https://doi.org/10.1016/0370-1573(88)90023-3}{\emph{Phys. Rept.}
  {\bfseries 168} (1988) 115}.

\bibitem{Hu:1993qa}
B.~L. Hu and A.~Matacz, \emph{{Quantum Brownian motion in a bath of parametric
  oscillators: A Model for system - field interactions}},
  \href{https://doi.org/10.1103/PhysRevD.49.6612}{\emph{Phys. Rev.} {\bfseries
  D49} (1994) 6612} [\href{https://arxiv.org/abs/gr-qc/9312035}{{\ttfamily
  gr-qc/9312035}}].

\bibitem{Hu:1986jj}
B.~L. Hu and H.~E. Kandrup, \emph{{Entropy Generation in Cosmological Particle
  Creation and Interactions: A Statistical Subdynamics Analysis}},
  \href{https://doi.org/10.1103/PhysRevD.35.1776}{\emph{Phys. Rev.} {\bfseries
  D35} (1987) 1776}.

\bibitem{Hsiang:2005pz}
J.-T. Hsiang and D.-S. Lee, \emph{{Influence on electron coherence from quantum
  electromagnetic fields in the presence of conducting plates}},
  \href{https://doi.org/10.1103/PhysRevD.73.065022}{\emph{Phys. Rev.}
  {\bfseries D73} (2006) 065022}
  [\href{https://arxiv.org/abs/hep-th/0512059}{{\ttfamily hep-th/0512059}}].

\bibitem{Hsiang:2007zb}
J.-T. Hsiang, T.-H. Wu and D.-S. Lee, \emph{{Stochastic Lorentz forces on a
  point charge moving near the conducting plate}},
  \href{https://doi.org/10.1103/PhysRevD.77.105021}{\emph{Phys. Rev.}
  {\bfseries D77} (2008) 105021}
  [\href{https://arxiv.org/abs/0706.3075}{{\ttfamily 0706.3075}}].

\bibitem{Kachru:2008yh}
S.~Kachru, X.~Liu and M.~Mulligan, \emph{{Gravity Duals of Lifshitz-like Fixed
  Points}}, \href{https://doi.org/10.1103/PhysRevD.78.106005}{\emph{Phys.Rev.}
  {\bfseries D78} (2008) 106005}
  [\href{https://arxiv.org/abs/0808.1725}{{\ttfamily 0808.1725}}].

\bibitem{Dong:2012se}
X.~Dong, S.~Harrison, S.~Kachru, G.~Torroba and H.~Wang, \emph{{Aspects of
  holography for theories with hyperscaling violation}},
  \href{https://doi.org/10.1007/JHEP06(2012)041}{\emph{JHEP} {\bfseries 1206}
  (2012) 041} [\href{https://arxiv.org/abs/1201.1905}{{\ttfamily 1201.1905}}].

\bibitem{Narayan:2012hk}
K.~Narayan, \emph{{On Lifshitz scaling and hyperscaling violation in string
  theory}}, \href{https://doi.org/10.1103/PhysRevD.85.106006}{\emph{Phys. Rev.}
  {\bfseries D85} (2012) 106006}
  [\href{https://arxiv.org/abs/1202.5935}{{\ttfamily 1202.5935}}].

\bibitem{Azeyanagi:2009pr}
T.~Azeyanagi, W.~Li and T.~Takayanagi, \emph{{On String Theory Duals of
  Lifshitz-like Fixed Points}},
  \href{https://doi.org/10.1088/1126-6708/2009/06/084}{\emph{JHEP} {\bfseries
  0906} (2009) 084} [\href{https://arxiv.org/abs/0905.0688}{{\ttfamily
  0905.0688}}].

\bibitem{Mateos:2011ix}
D.~Mateos and D.~Trancanelli, \emph{{The anisotropic N=4 super Yang-Mills
  plasma and its instabilities}},
  \href{https://doi.org/10.1103/PhysRevLett.107.101601}{\emph{Phys.Rev.Lett.}
  {\bfseries 107} (2011) 101601}
  [\href{https://arxiv.org/abs/1105.3472}{{\ttfamily 1105.3472}}].

\bibitem{Mateos:2011tv}
D.~Mateos and D.~Trancanelli, \emph{{Thermodynamics and Instabilities of a
  Strongly Coupled Anisotropic Plasma}},
  \href{https://doi.org/10.1007/JHEP07(2011)054}{\emph{JHEP} {\bfseries 1107}
  (2011) 054} [\href{https://arxiv.org/abs/1106.1637}{{\ttfamily 1106.1637}}].

\bibitem{Giataganas:2017koz}
D.~Giataganas, U.~Gursoy and J.~F. Pedraza, \emph{{Strongly-coupled anisotropic
  gauge theories and holography}},
  \href{https://arxiv.org/abs/1708.05691}{{\ttfamily 1708.05691}}.

\bibitem{Jain:2014vka}
S.~Jain, N.~Kundu, K.~Sen, A.~Sinha and S.~P. Trivedi, \emph{{A Strongly
  Coupled Anisotropic Fluid From Dilaton Driven Holography}},
  \href{https://doi.org/10.1007/JHEP01(2015)005}{\emph{JHEP} {\bfseries 01}
  (2015) 005} [\href{https://arxiv.org/abs/1406.4874}{{\ttfamily 1406.4874}}].

\bibitem{Donos:2016zpf}
A.~Donos, J.~P. Gauntlett and O.~Sosa-Rodriguez, \emph{{Anisotropic plasmas
  from axion and dilaton deformations}},
  \href{https://doi.org/10.1007/JHEP11(2016)002}{\emph{JHEP} {\bfseries 11}
  (2016) 002} [\href{https://arxiv.org/abs/1608.02970}{{\ttfamily
  1608.02970}}].

\bibitem{Edalati:2012tc}
M.~Edalati, J.~F. Pedraza and W.~Tangarife~Garcia, \emph{{Quantum Fluctuations
  in Holographic Theories with Hyperscaling Violation}},
  \href{https://doi.org/10.1103/PhysRevD.87.046001}{\emph{Phys. Rev.}
  {\bfseries D87} (2013) 046001}
  [\href{https://arxiv.org/abs/1210.6993}{{\ttfamily 1210.6993}}].

\bibitem{Giataganas:2014hma}
D.~Giataganas and K.~Sfetsos, \emph{{Non-integrability in non-relativistic
  theories}}, \href{https://doi.org/10.1007/JHEP06(2014)018}{\emph{JHEP}
  {\bfseries 06} (2014) 018} [\href{https://arxiv.org/abs/1403.2703}{{\ttfamily
  1403.2703}}].

\bibitem{Motl:2003cd}
L.~Motl and A.~Neitzke, \emph{{Asymptotic black hole quasinormal frequencies}},
  \href{https://doi.org/10.4310/ATMP.2003.v7.n2.a4}{\emph{Adv. Theor. Math.
  Phys.} {\bfseries 7} (2003) 307}
  [\href{https://arxiv.org/abs/hep-th/0301173}{{\ttfamily hep-th/0301173}}].

\bibitem{Maldacena:1996ix}
J.~M. Maldacena and A.~Strominger, \emph{{Black hole grey body factors and
  d-brane spectroscopy}},
  \href{https://doi.org/10.1103/PhysRevD.55.861}{\emph{Phys. Rev.} {\bfseries
  D55} (1997) 861} [\href{https://arxiv.org/abs/hep-th/9609026}{{\ttfamily
  hep-th/9609026}}].

\bibitem{Harmark:2007jy}
T.~Harmark, J.~Natario and R.~Schiappa, \emph{{Greybody Factors for
  d-Dimensional Black Holes}},
  \href{https://doi.org/10.4310/ATMP.2010.v14.n3.a1}{\emph{Adv. Theor. Math.
  Phys.} {\bfseries 14} (2010) 727}
  [\href{https://arxiv.org/abs/0708.0017}{{\ttfamily 0708.0017}}].

\bibitem{Giataganas:2015ksa}
D.~Giataganas, \emph{{$k$-Strings as Fundamental Strings}},
  \href{https://doi.org/10.1007/JHEP05(2015)134}{\emph{JHEP} {\bfseries 05}
  (2015) 134} [\href{https://arxiv.org/abs/1503.09180}{{\ttfamily
  1503.09180}}].

\bibitem{Giataganas:2013hwa}
D.~Giataganas and H.~Soltanpanahi, \emph{{Universal Properties of the Langevin
  Diffusion Coefficients}},
  \href{https://doi.org/10.1103/PhysRevD.89.026011}{\emph{Phys.Rev.} {\bfseries
  D89} (2014) 026011} [\href{https://arxiv.org/abs/1310.6725}{{\ttfamily
  1310.6725}}].

\bibitem{Giataganas:2013zaa}
D.~Giataganas and H.~Soltanpanahi, \emph{{Heavy Quark Diffusion in Strongly
  Coupled Anisotropic Plasmas}},
  \href{https://doi.org/10.1007/JHEP06(2014)047}{\emph{JHEP} {\bfseries 06}
  (2014) 047} [\href{https://arxiv.org/abs/1312.7474}{{\ttfamily 1312.7474}}].

\bibitem{Finazzo:2016mhm}
S.~I. Finazzo, R.~Critelli, R.~Rougemont and J.~Noronha, \emph{{Momentum
  transport in strongly coupled anisotropic plasmas in the presence of strong
  magnetic fields}}, \href{https://doi.org/10.1103/PhysRevD.94.054020,
  10.1103/PhysRevD.96.019903}{\emph{Phys. Rev.} {\bfseries D94} (2016) 054020}
  [\href{https://arxiv.org/abs/1605.06061}{{\ttfamily 1605.06061}}].

\bibitem{Herzog:2006gh}
C.~Herzog, A.~Karch, P.~Kovtun, C.~Kozcaz and L.~Yaffe, \emph{{Energy loss of a
  heavy quark moving through N=4 supersymmetric Yang-Mills plasma}},
  \href{https://doi.org/10.1088/1126-6708/2006/07/013}{\emph{JHEP} {\bfseries
  0607} (2006) 013} [\href{https://arxiv.org/abs/hep-th/0605158}{{\ttfamily
  hep-th/0605158}}].

\bibitem{Gubser:2006bz}
S.~S. Gubser, \emph{{Drag force in AdS/CFT}},
  \href{https://doi.org/10.1103/PhysRevD.74.126005}{\emph{Phys.Rev.} {\bfseries
  D74} (2006) 126005} [\href{https://arxiv.org/abs/hep-th/0605182}{{\ttfamily
  hep-th/0605182}}].

\bibitem{Gubser:2006nz}
S.~S. Gubser, \emph{{Momentum fluctuations of heavy quarks in the gauge-string
  duality}},
  \href{https://doi.org/10.1016/j.nuclphysb.2007.09.017}{\emph{Nucl.Phys.}
  {\bfseries B790} (2008) 175}
  [\href{https://arxiv.org/abs/hep-th/0612143}{{\ttfamily hep-th/0612143}}].

\bibitem{CasalderreySolana:2006rq}
J.~Casalderrey-Solana and D.~Teaney, \emph{{Heavy quark diffusion in strongly
  coupled N=4 Yang-Mills}},
  \href{https://doi.org/10.1103/PhysRevD.74.085012}{\emph{Phys.Rev.} {\bfseries
  D74} (2006) 085012} [\href{https://arxiv.org/abs/hep-ph/0605199}{{\ttfamily
  hep-ph/0605199}}].

\bibitem{CasalderreySolana:2007qw}
J.~Casalderrey-Solana and D.~Teaney, \emph{{Transverse Momentum Broadening of a
  Fast Quark in a N=4 Yang Mills Plasma}},
  \href{https://doi.org/10.1088/1126-6708/2007/04/039}{\emph{JHEP} {\bfseries
  0704} (2007) 039} [\href{https://arxiv.org/abs/hep-th/0701123}{{\ttfamily
  hep-th/0701123}}].

\bibitem{Giecold:2009cg}
G.~C. Giecold, E.~Iancu and A.~H. Mueller, \emph{{Stochastic trailing string
  and Langevin dynamics from AdS/CFT}},
  \href{https://doi.org/10.1088/1126-6708/2009/07/033}{\emph{JHEP} {\bfseries
  07} (2009) 033} [\href{https://arxiv.org/abs/0903.1840}{{\ttfamily
  0903.1840}}].

\bibitem{Gursoy:2009kk}
U.~Gursoy, E.~Kiritsis, G.~Michalogiorgakis and F.~Nitti, \emph{{Thermal
  Transport and Drag Force in Improved Holographic QCD}},
  \href{https://doi.org/10.1088/1126-6708/2009/12/056}{\emph{JHEP} {\bfseries
  0912} (2009) 056} [\href{https://arxiv.org/abs/0906.1890}{{\ttfamily
  0906.1890}}].

\bibitem{HoyosBadajoz:2009pv}
C.~Hoyos-Badajoz, \emph{{Drag and jet quenching of heavy quarks in a strongly
  coupled N=2* plasma}},
  \href{https://doi.org/10.1088/1126-6708/2009/09/068}{\emph{JHEP} {\bfseries
  0909} (2009) 068} [\href{https://arxiv.org/abs/0907.5036}{{\ttfamily
  0907.5036}}].

\bibitem{Gursoy:2010aa}
U.~Gursoy, E.~Kiritsis, L.~Mazzanti and F.~Nitti, \emph{{Langevin diffusion of
  heavy quarks in non-conformal holographic backgrounds}},
  \href{https://doi.org/10.1007/JHEP12(2010)088}{\emph{JHEP} {\bfseries 1012}
  (2010) 088} [\href{https://arxiv.org/abs/1006.3261}{{\ttfamily 1006.3261}}].

\bibitem{Kovtun:2004de}
P.~Kovtun, D.~T. Son and A.~O. Starinets, \emph{{Viscosity in strongly
  interacting quantum field theories from black hole physics}},
  \href{https://doi.org/10.1103/PhysRevLett.94.111601}{\emph{Phys. Rev. Lett.}
  {\bfseries 94} (2005) 111601}
  [\href{https://arxiv.org/abs/hep-th/0405231}{{\ttfamily hep-th/0405231}}].

\bibitem{Rebhan:2011vd}
A.~Rebhan and D.~Steineder, \emph{{Violation of the Holographic Viscosity Bound
  in a Strongly Coupled Anisotropic Plasma}},
  \href{https://doi.org/10.1103/PhysRevLett.108.021601}{\emph{Phys. Rev. Lett.}
  {\bfseries 108} (2012) 021601}
  [\href{https://arxiv.org/abs/1110.6825}{{\ttfamily 1110.6825}}].

\bibitem{Jain:2015txa}
S.~Jain, R.~Samanta and S.~P. Trivedi, \emph{{The Shear Viscosity in
  Anisotropic Phases}},
  \href{https://doi.org/10.1007/JHEP10(2015)028}{\emph{JHEP} {\bfseries 10}
  (2015) 028} [\href{https://arxiv.org/abs/1506.01899}{{\ttfamily
  1506.01899}}].

\bibitem{Erdmenger:2010xm}
J.~Erdmenger, P.~Kerner and H.~Zeller, \emph{{Non-universal shear viscosity
  from Einstein gravity}},
  \href{https://doi.org/10.1016/j.physletb.2011.04.009}{\emph{Phys. Lett.}
  {\bfseries B699} (2011) 301}
  [\href{https://arxiv.org/abs/1011.5912}{{\ttfamily 1011.5912}}].

\bibitem{Samanta:2016pic}
R.~Samanta, R.~Sharma and S.~P. Trivedi, \emph{{Shear viscosity in an
  anisotropic unitary Fermi gas}},
  \href{https://doi.org/10.1103/PhysRevA.96.053601}{\emph{Phys. Rev.}
  {\bfseries A96} (2017) 053601}
  [\href{https://arxiv.org/abs/1607.04799}{{\ttfamily 1607.04799}}].

\bibitem{Ge:2015owa}
X.-H. Ge, \emph{{Notes on shear viscosity bound violation in anisotropic
  models}}, \href{https://doi.org/10.1007/s11433-015-5776-2}{\emph{Sci. China
  Phys. Mech. Astron.} {\bfseries 59} (2016) 630401}
  [\href{https://arxiv.org/abs/1510.06861}{{\ttfamily 1510.06861}}].

\bibitem{Kolekar:2016pnr}
K.~S. Kolekar, D.~Mukherjee and K.~Narayan, \emph{{Hyperscaling violation and
  the shear diffusion constant}},
  \href{https://doi.org/10.1016/j.physletb.2016.06.046}{\emph{Phys. Lett.}
  {\bfseries B760} (2016) 86}
  [\href{https://arxiv.org/abs/1604.05092}{{\ttfamily 1604.05092}}].

\bibitem{Gubser:2006qh}
S.~S. Gubser, \emph{{Comparing the drag force on heavy quarks in N=4
  super-Yang-Mills theory and QCD}},
  \href{https://doi.org/10.1103/PhysRevD.76.126003}{\emph{Phys. Rev.}
  {\bfseries D76} (2007) 126003}
  [\href{https://arxiv.org/abs/hep-th/0611272}{{\ttfamily hep-th/0611272}}].

\bibitem{CasalderreySolana:2009rm}
J.~Casalderrey-Solana, K.-Y. Kim and D.~Teaney, \emph{{Stochastic String Motion
  Above and Below the World Sheet Horizon}},
  \href{https://doi.org/10.1088/1126-6708/2009/12/066}{\emph{JHEP} {\bfseries
  12} (2009) 066} [\href{https://arxiv.org/abs/0908.1470}{{\ttfamily
  0908.1470}}].

\bibitem{Rajagopal:2015roa}
K.~Rajagopal and A.~V. Sadofyev, \emph{{Chiral drag force}},
  \href{https://doi.org/10.1007/JHEP10(2015)018}{\emph{JHEP} {\bfseries 10}
  (2015) 018} [\href{https://arxiv.org/abs/1505.07379}{{\ttfamily
  1505.07379}}].

\bibitem{Herzog:2007kh}
C.~P. Herzog and A.~Vuorinen, \emph{{Spinning Dragging Strings}},
  \href{https://doi.org/10.1088/1126-6708/2007/10/087}{\emph{JHEP} {\bfseries
  10} (2007) 087} [\href{https://arxiv.org/abs/0708.0609}{{\ttfamily
  0708.0609}}].

\bibitem{Akamatsu:2015kaa}
Y.~Akamatsu, \emph{{Langevin dynamics and decoherence of heavy quarks at high
  temperatures}}, \href{https://doi.org/10.1103/PhysRevC.92.044911}{\emph{Phys.
  Rev.} {\bfseries C92} (2015) 044911}
  [\href{https://arxiv.org/abs/1503.08110}{{\ttfamily 1503.08110}}].

\bibitem{Horowitz:2015dta}
W.~A. Horowitz, \emph{{Fluctuating heavy quark energy loss in a strongly
  coupled quark-gluon plasma}},
  \href{https://doi.org/10.1103/PhysRevD.91.085019}{\emph{Phys. Rev.}
  {\bfseries D91} (2015) 085019}
  [\href{https://arxiv.org/abs/1501.04693}{{\ttfamily 1501.04693}}].

\bibitem{Nakamura:2013yqa}
S.~Nakamura and H.~Ooguri, \emph{{Out of Equilibrium Temperature from
  Holography}}, \href{https://doi.org/10.1103/PhysRevD.88.126003}{\emph{Phys.
  Rev.} {\bfseries D88} (2013) 126003}
  [\href{https://arxiv.org/abs/1309.4089}{{\ttfamily 1309.4089}}].

\bibitem{Arean:2016het}
D.~Arean, L.~A. Pando~Zayas, L.~Patino and M.~Villasante, \emph{{Velocity
  Statistics in Holographic Fluids: Magnetized Quark-Gluon Plasma and
  Superfluid Flow}}, \href{https://doi.org/10.1007/JHEP10(2016)158}{\emph{JHEP}
  {\bfseries 10} (2016) 158}
  [\href{https://arxiv.org/abs/1606.03068}{{\ttfamily 1606.03068}}].

\bibitem{Roy:2009sw}
S.~Roy, \emph{{Holography and drag force in thermal plasma of non-commutative
  Yang-Mills theories in diverse dimensions}},
  \href{https://doi.org/10.1016/j.physletb.2009.10.095}{\emph{Phys. Lett.}
  {\bfseries B682} (2009) 93}
  [\href{https://arxiv.org/abs/0907.0333}{{\ttfamily 0907.0333}}].

\bibitem{Horowitz:2009pw}
W.~A. Horowitz and Y.~V. Kovchegov, \emph{{Shock Treatment: Heavy Quark Drag in
  a Novel AdS Geometry}},
  \href{https://doi.org/10.1016/j.physletb.2009.07.077}{\emph{Phys. Lett.}
  {\bfseries B680} (2009) 56}
  [\href{https://arxiv.org/abs/0904.2536}{{\ttfamily 0904.2536}}].

\bibitem{Caceres:2006as}
E.~Caceres and A.~Guijosa, \emph{{On Drag Forces and Jet Quenching in Strongly
  Coupled Plasmas}},
  \href{https://doi.org/10.1088/1126-6708/2006/12/068}{\emph{JHEP} {\bfseries
  12} (2006) 068} [\href{https://arxiv.org/abs/hep-th/0606134}{{\ttfamily
  hep-th/0606134}}].

\bibitem{Ahmadvand:2015gfi}
M.~Ahmadvand and K.~Bitaghsir~Fadafan, \emph{{Energy loss at zero temperature
  from extremal black holes}},
  \href{https://arxiv.org/abs/1512.05290}{{\ttfamily 1512.05290}}.

\bibitem{Zhang:2018mqt}
Z.-q. Zhang, K.~Ma and D.-f. Hou, \emph{{Drag force in strongly coupled
  supersymmetric Yang–Mills plasma in a magnetic field}},
  \href{https://doi.org/10.1088/1361-6471/aaa097}{\emph{J. Phys.} {\bfseries
  G45} (2018) 025003} [\href{https://arxiv.org/abs/1802.01912}{{\ttfamily
  1802.01912}}].

\bibitem{Hartnoll:2009ns}
S.~A. Hartnoll, J.~Polchinski, E.~Silverstein and D.~Tong, \emph{{Towards
  strange metallic holography}},
  \href{https://doi.org/10.1007/JHEP04(2010)120}{\emph{JHEP} {\bfseries 04}
  (2010) 120} [\href{https://arxiv.org/abs/0912.1061}{{\ttfamily 0912.1061}}].

\bibitem{Kiritsis:2012ta}
E.~Kiritsis, \emph{{Lorentz violation, Gravity, Dissipation and Holography}},
  \href{https://doi.org/10.1007/JHEP01(2013)030}{\emph{JHEP} {\bfseries 01}
  (2013) 030} [\href{https://arxiv.org/abs/1207.2325}{{\ttfamily 1207.2325}}].

\bibitem{Fadafan:2009an}
K.~B. Fadafan, \emph{{Drag force in asymptotically Lifshitz spacetimes}},
  \href{https://arxiv.org/abs/0912.4873}{{\ttfamily 0912.4873}}.

\bibitem{CasalderreySolana:2011us}
J.~Casalderrey-Solana, H.~Liu, D.~Mateos, K.~Rajagopal and U.~A. Wiedemann,
  \emph{{Gauge/String Duality, Hot QCD and Heavy Ion Collisions}},
  \href{https://arxiv.org/abs/1101.0618}{{\ttfamily 1101.0618}}.

\bibitem{Giataganas:2012zy}
D.~Giataganas, \emph{{Probing strongly coupled anisotropic plasma}},
  \href{https://doi.org/10.1007/JHEP07(2012)031}{\emph{JHEP} {\bfseries 1207}
  (2012) 031} [\href{https://arxiv.org/abs/1202.4436}{{\ttfamily 1202.4436}}].

\bibitem{Giataganas:2013lga}
D.~Giataganas, \emph{{Observables in Strongly Coupled Anisotropic Theories}},
  {\emph{PoS} {\bfseries Corfu2012} (2013) 122}
  [\href{https://arxiv.org/abs/1306.1404}{{\ttfamily 1306.1404}}].

\bibitem{Gubser:2007ga}
S.~S. Gubser, S.~S. Pufu and A.~Yarom, \emph{{Sonic booms and diffusion wakes
  generated by a heavy quark in thermal AdS/CFT}},
  \href{https://doi.org/10.1103/PhysRevLett.100.012301}{\emph{Phys. Rev. Lett.}
  {\bfseries 100} (2008) 012301}
  [\href{https://arxiv.org/abs/0706.4307}{{\ttfamily 0706.4307}}].

\bibitem{Gubser:2007ni}
S.~S. Gubser and A.~Yarom, \emph{{Universality of the diffusion wake in the
  gauge-string duality}},
  \href{https://doi.org/10.1103/PhysRevD.77.066007}{\emph{Phys. Rev.}
  {\bfseries D77} (2008) 066007}
  [\href{https://arxiv.org/abs/0709.1089}{{\ttfamily 0709.1089}}].

\bibitem{CasalderreySolana:2006sq}
J.~Casalderrey-Solana, E.~V. Shuryak and D.~Teaney, \emph{{Hydrodynamic flow
  from fast particles}},
  \href{https://arxiv.org/abs/hep-ph/0602183}{{\ttfamily hep-ph/0602183}}.

\bibitem{Friess:2006fk}
J.~J. Friess, S.~S. Gubser, G.~Michalogiorgakis and S.~S. Pufu, \emph{{The
  Stress tensor of a quark moving through N=4 thermal plasma}},
  \href{https://doi.org/10.1103/PhysRevD.75.106003}{\emph{Phys. Rev.}
  {\bfseries D75} (2007) 106003}
  [\href{https://arxiv.org/abs/hep-th/0607022}{{\ttfamily hep-th/0607022}}].

\bibitem{Gubser:2009sn}
S.~S. Gubser, S.~S. Pufu, F.~D. Rocha and A.~Yarom, \emph{{Energy loss in a
  strongly coupled thermal medium and the gauge-string duality}},  in
  \emph{Quark-gluon plasma 4} (R.~C. Hwa and X.-N. Wang, eds.), pp.~1--59.
\newblock 2010.
\newblock \href{https://arxiv.org/abs/0902.4041}{{\ttfamily 0902.4041}}.
\newblock \href{https://doi.org/10.1142/9789814293297_0001}{DOI}.

\bibitem{bgflattice}
G.~Backus and F.~Gilbert, \emph{The resolving power of gross earth data},
  \href{https://doi.org/10.1111/j.1365-246X.1968.tb00216.x}{\emph{Geophysical
  Journal International} {\bfseries 16} (1968) 169}.

\bibitem{Francis:2015daa}
A.~Francis, O.~Kaczmarek, M.~Laine, T.~Neuhaus and H.~Ohno,
  \emph{{Nonperturbative estimate of the heavy quark momentum diffusion
  coefficient}}, \href{https://doi.org/10.1103/PhysRevD.92.116003}{\emph{Phys.
  Rev.} {\bfseries D92} (2015) 116003}
  [\href{https://arxiv.org/abs/1508.04543}{{\ttfamily 1508.04543}}].

\bibitem{CaronHuot:2007gq}
S.~Caron-Huot and G.~D. Moore, \emph{{Heavy quark diffusion in perturbative QCD
  at next-to-leading order}},
  \href{https://doi.org/10.1103/PhysRevLett.100.052301}{\emph{Phys. Rev. Lett.}
  {\bfseries 100} (2008) 052301}
  [\href{https://arxiv.org/abs/0708.4232}{{\ttfamily 0708.4232}}].

\bibitem{Ding:2010ga}
H.~T. Ding, A.~Francis, O.~Kaczmarek, F.~Karsch, E.~Laermann and W.~Soeldner,
  \emph{{Thermal dilepton rate and electrical conductivity: An analysis of
  vector current correlation functions in quenched lattice QCD}},
  \href{https://doi.org/10.1103/PhysRevD.83.034504}{\emph{Phys. Rev.}
  {\bfseries D83} (2011) 034504}
  [\href{https://arxiv.org/abs/1012.4963}{{\ttfamily 1012.4963}}].

\bibitem{Burnier:2012ts}
Y.~Burnier and M.~Laine, \emph{{Towards flavour diffusion coefficient and
  electrical conductivity without ultraviolet contamination}},
  \href{https://doi.org/10.1140/epjc/s10052-012-1902-8}{\emph{Eur. Phys. J.}
  {\bfseries C72} (2012) 1902}
  [\href{https://arxiv.org/abs/1201.1994}{{\ttfamily 1201.1994}}].

\bibitem{Karch:2002sh}
A.~Karch and E.~Katz, \emph{{Adding flavor to AdS / CFT}},
  \href{https://doi.org/10.1088/1126-6708/2002/06/043}{\emph{JHEP} {\bfseries
  06} (2002) 043} [\href{https://arxiv.org/abs/hep-th/0205236}{{\ttfamily
  hep-th/0205236}}].

\bibitem{Erdmenger:2007cm}
J.~Erdmenger, N.~Evans, I.~Kirsch and E.~Threlfall, \emph{{Mesons in
  Gauge/Gravity Duals - A Review}},
  \href{https://doi.org/10.1140/epja/i2007-10540-1}{\emph{Eur. Phys. J.}
  {\bfseries A35} (2008) 81} [\href{https://arxiv.org/abs/0711.4467}{{\ttfamily
  0711.4467}}].

\bibitem{Burrington:2004id}
B.~A. Burrington, J.~T. Liu, L.~A. Pando~Zayas and D.~Vaman, \emph{{Holographic
  duals of flavored N=1 super Yang-mills: Beyond the probe approximation}},
  \href{https://doi.org/10.1088/1126-6708/2005/02/022}{\emph{JHEP} {\bfseries
  02} (2005) 022} [\href{https://arxiv.org/abs/hep-th/0406207}{{\ttfamily
  hep-th/0406207}}].

\bibitem{Nunez:2010sf}
C.~Nunez, A.~Paredes and A.~V. Ramallo, \emph{{Unquenched Flavor in the
  Gauge/Gravity Correspondence}},
  \href{https://doi.org/10.1155/2010/196714}{\emph{Adv. High Energy Phys.}
  {\bfseries 2010} (2010) 196714}
  [\href{https://arxiv.org/abs/1002.1088}{{\ttfamily 1002.1088}}].

\bibitem{Kirsch:2005uy}
I.~Kirsch and D.~Vaman, \emph{{The D3 / D7 background and flavor dependence of
  Regge trajectories}},
  \href{https://doi.org/10.1103/PhysRevD.72.026007}{\emph{Phys. Rev.}
  {\bfseries D72} (2005) 026007}
  [\href{https://arxiv.org/abs/hep-th/0505164}{{\ttfamily hep-th/0505164}}].

\bibitem{Giataganas:2011nz}
D.~Giataganas and N.~Irges, \emph{{Flavor Corrections in the Static Potential
  in Holographic QCD}},
  \href{https://doi.org/10.1103/PhysRevD.85.046001}{\emph{Phys.Rev.} {\bfseries
  D85} (2012) 046001} [\href{https://arxiv.org/abs/1104.1623}{{\ttfamily
  1104.1623}}].

\bibitem{Bigazzi:2014qsa}
F.~Bigazzi and A.~L. Cotrone, \emph{{Holographic QCD with Dynamical Flavors}},
  \href{https://doi.org/10.1007/JHEP01(2015)104}{\emph{JHEP} {\bfseries 01}
  (2015) 104} [\href{https://arxiv.org/abs/1410.2443}{{\ttfamily 1410.2443}}].

\bibitem{Alho:2012mh}
T.~Alho, M.~Jarvinen, K.~Kajantie, E.~Kiritsis and K.~Tuominen, \emph{{On
  finite-temperature holographic QCD in the Veneziano limit}},
  \href{https://doi.org/10.1007/JHEP01(2013)093}{\emph{JHEP} {\bfseries 01}
  (2013) 093} [\href{https://arxiv.org/abs/1210.4516}{{\ttfamily 1210.4516}}].

\bibitem{Faedo:2017aoe}
A.~F. Faedo, D.~Mateos, C.~Pantelidou and J.~Tarrio, \emph{{Towards a
  Holographic Quark Matter Crystal}},
  \href{https://doi.org/10.1007/JHEP10(2017)139}{\emph{JHEP} {\bfseries 10}
  (2017) 139} [\href{https://arxiv.org/abs/1707.06989}{{\ttfamily
  1707.06989}}].

\bibitem{Li:2016gtz}
S.-w. Li and T.~Jia, \emph{{Dynamically flavored description of holographic QCD
  in the presence of a magnetic field}},
  \href{https://doi.org/10.1103/PhysRevD.96.066032}{\emph{Phys. Rev.}
  {\bfseries D96} (2017) 066032}
  [\href{https://arxiv.org/abs/1604.07197}{{\ttfamily 1604.07197}}].

\end{thebibliography}

\end{document}